 \documentstyle[preprint,aps]{revtex}
 \tightenlines
  \newcommand{\beq}{\begin{equation}}
  \newcommand{\eeq}{\end{equation}}
  \newcommand{\beql}[1]{\begin{equation}\label{eq:#1}}
  \newcommand{\beqa}{\begin{eqnarray}}
  \newcommand{\eeqa}{\end{eqnarray}}
  \newcommand{\beqas}{\begin{eqnarray*}}
  \newcommand{\eeqas}{\end{eqnarray*}}
   \newcommand{\qed}{$\Box$}
  \newcommand{\R}{{\bf R}}
  \newcommand{\bA}{{\bf A}}
  \newcommand{\bL}{{\bf L}}
  \newcommand{\bP}{{\bf P}}
  \newcommand{\bS}{{\bf S}}
  \newcommand{\bT}{{\bf T}}
  \newcommand{\bX}{{\bf X}}
  \newcommand{\cB}{{\cal B}}
  \newcommand{\cC}{{\cal C}}
  \newcommand{\cD}{{\cal D}}
  \newcommand{\cE}{{\cal E}}
  \newcommand{\cH}{{\cal H}}
  \newcommand{\cK}{{\cal K}}
  \newcommand{\cL}{{\cal L}}

  \newcommand{\da}{\dagger}
  
  \newcommand{\ep}{\epsilon}
  \newcommand{\et}{\eta}

  \newcommand{\la}{\lambda}
  \newcommand{\mb}{\mbox}
  \newcommand{\nn}{\nonumber}
  
  \newcommand{\ph}{\phi}
  \newcommand{\ps}{\psi}
  \newcommand{\rh}{\rho}
  \newcommand{\si}{\sigma}
  \newcommand{\ta}{\tau}
  \newcommand{\th}{\theta}

  \newcommand{\tc}{\tau c}
  \newcommand{\De}{\Delta}

  \newcommand{\Ph}{\Phi}                                            
  \newcommand{\Ps}{\Psi}

  \newcommand{\Tr}{\mbox{\rm Tr}}
                                          
  \newcommand{\ba}{{\bf a}}
  \newcommand{\bb}{{\bf b}}
  \newcommand{\bc}{{\bf c}}
  \newcommand{\bm}{{\bf m}}
  \newcommand{\bx}{{\bf x}}
  
  \newcommand{\eq}[1]{(\ref{eq:#1})}

\newcommand{\bra}[1]{\langle#1|}
\newcommand{\ket}[1]{|#1\rangle}
\newcommand{\braket}[1]{\langle#1\rangle}
\newcommand{\bracket}[1]{\langle#1\rangle}
\renewcommand{\AA}{\bA(\ba)}

\begin{document}
\draft
\title{Operations, Disturbance, and Simultaneous Measurability}
\author{Masanao Ozawa}
\address{School of Informatics and Sciences,
Nagoya University, Chikusa-ku, Nagoya 464-8601, Japan}
\maketitle
\begin{abstract}
Quantum mechanics predicts the joint probability distributions of the 
outcomes of simultaneous measurements of commuting observables,
but the current formulation lacks the operational definition of
simultaneous measurements.
In order to provide foundations of joint statistics of local general 
measurements on entangled systems in a general theoretical
framework, the question is answered as to under what condition the 
outputs of two measuring apparatuses satisfy the joint probability 
formula for simultaneous measurements of their observables.
For this purpose, all the possible state changes caused by
measurements of an observable are characterized and the notion 
of disturbance in measurement is formalized in terms of operations 
derived by the measuring interaction.
\end{abstract}
\narrowtext
\section{Introduction}

The probability distribution of the outcome of a measurement
is determined by the observable to be measured and the state 
at the time of the measurement, but
the joint probability distribution of the outcomes of
two successive measurements on the same object depends on 
how the first measurement disturbs the object.
The disturbance depends not only on the observable and the state 
but also the apparatus to be used.
Thus, the joint probability distribution of successive measurements 
will be closely related to how the apparatus disturbs the object.
It would be an interesting and significant problem to investigate
the relation between the disturbance and the joint probability
distribution, although there has been no systematic approach
to the problem.
This paper investigates in particular the relation between the 
disturbance and the joint probability formula for simultaneous 
measurements.

In quantum mechanics observables are represented by linear operators, 
for which the product operation is not necessarily commutative.
Two observables are represented by commuting operators if and only if
they are simultaneously measurable,
and then quantum mechanics predicts the joint probability distribution 
of the outcomes of their simultaneous measurement 
\cite[p.~228]{vN55}.
But, it has not been answered fully what measurement can be considered
as a simultaneous measurement of those observables. 

The current formulation has two arguments to show how commuting 
observables can be measured simultaneously.
The first argument is based on the fact that any commuting
observables $A$ and $B$ have the third observable $C$ for which $A$ and $B$
are functions of $C$ \cite[p.~173]{vN55}.  
In this case, the measurement of $C$ gives also 
the outcomes of the $A$-measurement and the $B$-measurement simultaneously
\cite[p.~228]{vN55}.
This argument gives one special instance of simultaneous measurement,
but it is quite open from
this argument how a pair of measuring apparatuses for 
$A$ and $B$ makes a simultaneous measurement of $A$ and $B$.

The second argument assumes the projection postulate proposed 
by L\"{u}ders \cite{Lud51}.
The projection postulate determines uniquely the state after the measurement
conditional upon the outcome of the measurement, so that
for the successive measurement of any pair of discrete observables 
$A$ and $B$, the joint probability distribution of their outcomes
is determined.
According to this probability distribution,
if $A$ and $B$ commute, we have the standard joint probability
formula for the simultaneous measurement of $A$ and $B$.
Thus, under the projection postulate, 
the successive measurements of $A$ and $B$ are considered 
effectively as their simultaneous measurement.

If we would restrict the class of measurements to those satisfying 
the projection postulate, any successive measurements of commuting 
observables could be considered as a simultaneous measurement.
This approach has, however, the following limitations:
(i) Some of the most familiar measuring apparatuses such as photon 
counters do not satisfy the projection postulate.
(ii) When the observable has continuous spectrum, no measurements
satisfy the repeatability hypothesis \cite{84QC,85CA}, 
so that the projection postulate cannot be formulated properly 
for measurements of continuous observables.
(iii) The measurement of a function of an observable $C$ such as
$A=f(C)$ using the apparatus measuring $C$ does not satisfy 
the projection postulate in general \cite{Lud51}.

In fact, Einstein, Podolsky, and Rosen (EPR) \cite{EPR35} have
derived the joint probability formula for the outcomes of 
measurements of two observables pertaining to entangled subsystems 
{\em based on} the projection postulate.
This EPR-correlation is indeed one instance of the joint probability 
formula for the simultaneous measurement.
The EPR-correlation has been experimentally tested by optical 
experiments \cite{AGR82,ADR82}.
However, those optical experiments use the photon counting and violate
EPR's original assumption that the measurements satisfy the projection 
postulate.
The recent realizations of quantum teleportation \cite{BPMEWZ97,BBMHP97}
are also optical realizations of the EPR-correlation that violates EPR's 
assumption of the projection postulate.
Thus, if we should be restricted to measurements satisfying the projection 
postulate, the scope of measurement theory would exclude most of 
the recent results in quantum information processing 
using entanglement \cite{BD00}.

\sloppy
The modern measurement theory 
\cite{84QC,85CA,Dav76,Kra83,85CC,86IQ,89RS,BLM91} extends
the scope from the measurements satisfying the projection postulate 
to more general measurements described by operations and effects 
or more generally by operation valued measures introduced from 
an axiomatic motivation \cite{HK64,Lud67,DL70}.
Thus, it is natural to expect to have a well-defined way to 
calculate probabilities in any combinations of measuring apparatuses,
once we have identified the operation valued measure in the 
general theory with the given model of measuring apparatus.
However, the determination of the operation valued measure
so far has relied on the projection postulate \cite{Kra83} 
or the joint probability formula \cite{84QC,83CR}.
Thus, the foundations of the modern approach in the present status
involve the same difficulty as establishing the joint probability 
formula without assuming the projection postulate.

In this paper we shall abandon the projection postulate 
as a universal quantum rule and consider the following problem:
{\em under what condition can a successive measurement
of two or more observables be considered as a simultaneous
measurement of those observables?}

The prospective solution could be stated in the intuitive language
that the preceding measurement does not disturb the observable to be 
measured later. 
However, in the current quantum mechanics very little
has been known about the disturbance caused by general measurement
beyond the projection postulate. 
In order to answer the question in the rigorous language, 
this paper will attempt to develop a theory of
disturbance in general measurements as well as to determine the possible
state changes caused by general measurements of observables.
The justification of joint probability formula and the EPR-correlation 
will then follow without assuming the projection postulate.

Section II defines the simultaneous measurement for a pair of 
measuring apparatus.  
Sections III and IV discuss simultaneous measurements under 
the repeatability hypothesis and the projection postulate with
indicating their limitations.
The following three sections develop theory of general measurement.
Section V introduces the nonselective operations and their duals.
Section VI discusses the Davies-Lewis postulate for the existence of
operation valued measures corresponding to apparatuses and shows that
the two justifications of their postulate known so far involve
the same difficulty as the one in establishing the joint probability 
formula without assuming the projection postulate.
Section VII gives a new justification of the Davies-Lewis postulate
without assuming the projection postulate or the joint probability formula
and proves the factoring property of operation valued measures.
Section VIII formulates the disturbance in measurement and establishes
in the rigorous language the relation between the disturbance and
the joint probability formula.
Sections IX and X apply the above result to the EPR-correlation
and the minimum disturbing measurement.
Section XI concludes the paper with some remarks on the uncertainty 
principle.

\section{Statistical formula for simultaneous measurements}
\label{se:1229a}

\subsection{The Born statistical formula}

To formulate the problem precisely, let ${\bf S}$ be a quantum system 
with the Hilbert space ${\cal H}$ of state vectors.
We shall distinguish measuring apparatuses by their own output 
variables \cite{00MN}, denoting by $\bA(\bx)$ the apparatus measuring 
the system $\bS$ with the output variable $\bx$, which, we assume, 
takes values in the real line $\R$.
We shall denote by ``$\bx(t)\in\De$'' the probabilistic event
that the outcome of the measurement using the apparatus $\bA(\bx)$ 
at the time $t$ is in a Borel set $\De$ in the real line $\R$.  
(Throughout this paper, ``Borel set'' can be replaced by 
``interval'' for simplifying the presentation without any loss of
generality.)

Let $A$ be an observable of ${\bf S}$. 
The spectral projection of $A$ corresponding 
to a Borel set $\Delta$ is denoted by $E^{A}(\Delta)$.
According to the Born statistical formula, an apparatus $\bA(\ba)$ 
with output variable $\ba$ is said to {\em measure} 
an observable $A$ at the time $t$ if the relation
\beql{B1}\label{eq:0328b}
\Pr\{\ba(t)\in\De\}=\Tr[E^{A}(\De)\rh(t)]
\eeq
holds for the state $\rh(t)$ of the system $\bS$ at the time $t$.  
The state $\rh(t)$ is called the {\em input state} to the 
apparatus $\bA(\ba)$ and is taken to be an arbitrary density
operator.

The relation between the present formulation based on spectral
projections due to von Neumann \cite{vN55} 
and Dirac's formulation \cite{Dir58} is as follows.
If the observable $A$ has the Dirac type spectral decomposition
$$
A=\sum_{\nu}\sum_{\mu}\mu\ket{\mu,\nu}\bra{\mu,\nu}
+\sum_{\nu}\int\lambda\ket{\lambda,\nu}\bra{\lambda,\nu}\,d\lambda,
$$
where $\mu$ varies over the discrete eigenvalues, $\lambda$ varies over
the continuous eigenvalues, and $\nu$ is the degeneracy parameter,
then we have
$$
E^{A}(\Delta)=\sum_{\nu}\sum_{\mu\in\Delta}\ket{\mu,\nu}\bra{\mu,\nu}
+\sum_{\nu}\int_{\Delta}\ket{\lambda,\nu}\bra{\lambda,\nu}\,d\lambda.
$$
In this case, we have
\beqas
\lefteqn{\Tr[E^{A}(\De)\rh(t)]}\quad\\
&=&\sum_{\nu}\sum_{\mu\in\De}\bracket{\mu,\nu|\rh(t)|\mu,\nu}
+\sum_{\nu}\int_{\De}
\bracket{\lambda,\nu|\rh(t)|\lambda,\nu}\,d\lambda.
\eeqas

\subsection{Simultaneous measurements using one apparatus}

Any commuting observables $A$ and $B$ are simultaneously measurable 
and the joint probability distribution of the outcomes of their 
simultaneous measurement is given by
\beql{13}
\Pr\{\ba(t)\in\De,\bb(t)\in\De'\}
=\Tr[E^{A}(\De)E^{B}(\De')\rh(t)],
\eeq
where $\De$ and $\De'$ are arbitrary Borel sets and 
$\ba$ and $\bb$ denote the output variables of the apparatuses
measuring $A$ and $B$ at the time $t$, respectively.

A well-known proof of this formula from \eq{0328b}
runs as follows \cite[p.~228]{vN55}. 
Since $A$ and $B$ are commutable, 
there exist an observable $C$ and real-valued functions
$f$ and $g$ such that $A=f(C)$ and $B=g(C)$ \cite[p.~173]{vN55}.
Their spectral projections satisfy the relations
\beqas
E^{A}(\De)&=&E^{C}(f^{-1}(\De)),\\
E^{B}(\De')&=&E^{C}(g^{-1}(\De')).
\eeqas 
For the outcome $c$ of the $C$-measurement, one defines the outcome of the
$A$-measurement to be $f(c)$ and the outcome of the $B$-measurement to be
$g(c)$. Let $\ba$, $\bb$, and $\bc$ be the output variables of 
the measurements of $A$, $B$, and $C$, respectively.
Then, we have
\beqas
\lefteqn{\Pr\{\ba(t)\in\De,\bb(t)\in\De'\}}\quad\\
&=&\Pr\{\bc(t)\in f^{-1}(\De),\bc(t)\in g^{-1}(\De')\}\\
&=&\Pr\{\bc(t)\in f^{-1}(\De)\cap g^{-1}(\De')\}\\
&=&\Tr[E^{C}(f^{-1}(\De)\cap g^{-1}(\De'))\rh(t)]\\
&=&\Tr[E^{C}(f^{-1}(\De))E^{C}(g^{-1}(\De'))\rh(t)]\\
&=&\Tr[E^{A}(\De)E^{B}(\De')\rh(t)].
\eeqas 
Thus, their outcomes satisfy \eq{13} so that the measurement of $C$ at
the time $t$ gives a simultaneous measurement of $A$ and $B$.  

\subsection{Simultaneous measurements using two apparatuses}

The above proof gives one special instance of simultaneous measurement
which uses one measuring apparatus with two output variables, 
but it is rather open when a pair of measuring apparatuses for 
$A$ and $B$ makes a simultaneous measurement of $A$ and $B$.

In order to formulate this problem precisely, suppose that the 
observer measures $A$ at the time $t$ using the apparatus $\bA(\ba)$.
Let $t+\De t$ be the time just after the $\bA(\ba)$-measurement. 
This means precisely that $t+\De t$ is the instant of the time 
just after the interaction is turned off between $\bA(\ba)$ 
and $\bS$ and that after the time $t+\De t$ the object $\bS$ 
is free from the apparatus $\bA(\ba)$.  (Note that the last 
condition precludes the recoupling of the system with the 
apparatus.)

Let $\bA(\bb)$ be another apparatus measuring an 
observable $B$ of $\bS$ with output variable $\bb$.
If the measurement using $\bA(\bb)$ is turned on at the time $t+\De t$, 
the two measurements are called the {\em successive measurement} 
using $\bA(\ba)$ and $\bA(\bb)$ (in this order).
Then, {\em the successive measurement using $\bA(\ba)$ and $\bA(\bb)$ 
is defined to be a simultaneous measurement of $A$ and $B$ if and only
if the joint probability distribution of their output variables
$\ba$ and $\bb$ satisfies the standard joint probability formula}
\beql{0328c}
\Pr\{\ba(t)\in\De,\bb(t+\De t)\in\De'\}=
\Tr[E^{A}(\De)E^{B}(\De')\rh(t)].\label{eq:12}
\eeq

It should be noted that the validity of the above relation
depends on how the apparatus $\bA(\ba)$ disturbs $\bS$ 
during the first measurement but does not depend on the property
of the apparatus $\bA(\bb)$ as long as $\bA(\bb)$ 
measures the observable $B$ in any input state.
Therefore, the problem to be considered is to find the necessary and
sufficient condition on the apparatus $\bA(\ba)$ measuring $A$
in order for the successive measurement using $\bA(\ba)$ and an
arbitrary apparatus $\bA(\bb)$ measuring $B$ to satisfy \eq{0328c}.

In the conventional approach, von Neumann \cite[p.~224]{vN55} proved 
that if two observable are simultaneously measurable, then they are 
represented by commuting operators under the repeatability 
hypothesis.  (The repeatability hypothesis will be discussed in detail
in the next section.)
The following theorem, though an easy consequence of the 
definition, shows that the simultaneous measurability extended 
by the above definition is still consistent with the old one.

{\bf Theorem 1.}
{\em If the successive measurement of $A$ and $B$ 
using apparatuses $\bA(\ba)$ and $\bA(\bb)$, respectively, 
is a simultaneous measurement of observables $A$ and $B$,
then $A$ and $B$ commute.
}

{\em Proof.}
Suppose that (\ref{eq:12}) holds.
By the positivity of probability, the both sides are nonnegative.
Since $\rho(t)$ is arbitrary, the product
$E^{A}(\Delta)E^{B}(\Delta')$ is a positive self-adjoint operator so that
$E^{A}(\Delta)$ and $E^{B}(\Delta')$ commute.
Since $\Delta$ and $\Delta'$ are arbitrary, it follows that
$A$ and $B$ commute. 
\qed 

In the following two sections we shall re-examine the conventional 
approach from the operational point of view before starting with 
the general considerations.

\section{Simultaneous measurements under the repeatability hypothesis}

\subsection{Von Neumann's formulation}

The conventional approach to measurement theory supposes 
that the measurement leaves the measured system
in the eigenstate corresponding to the outcome 
of the measurement \cite{vN55,Sch35}. 
This assumption is equivalent to the {\em repeatability
hypothesis} formulated by von Neumann \cite[p.~335]{vN55} as follows.

(M) {\em If a physical quantity is measured twice in succession in a
system, then we get the same value each time.}

Even though the repeatability hypothesis has been posed as a universal 
law by von Neumann \cite[p.~213]{vN55} based on the experiment 
due to Compton and Simmons, in the modern approach
it characterizes merely a class of measuring apparatuses.
Thus, in what follows, {\em by saying that the apparatus $\bA(\ba)$ 
satisfies the repeatability hypothesis it is meant precisely 
that the repeatability hypothesis holds for the repeated measurement of $A$ 
using the apparatus $\bA(\ba)$ for the first $A$-measurement.}

\subsection{Repeatability hypothesis and joint probability}

Suppose that the system $\bS$ is measured at the time $t$
by an apparatus $\bA(\ba)$.
Let $t+\De t$ be the time just after the measurement.
For any Borel set $\Delta$, 
let $\rho(t+\Delta t|\ba(t)\in\Delta)$ be the state at $t+\Delta t$ 
of ${\bf S}$ conditional upon $\ba(t)\in\Delta$.
Thus, if the system ${\bf S}$ is sampled randomly from the subensemble 
of the similar systems that yield the outcome of the 
$\bA(\ba)$-measurement in the Borel set $\Delta$, then ${\bf S}$ is 
in the state $\rho(t+\Delta t|\ba(t)\in\Delta)$ at the time $t+\Delta t$.
When $\Pr\{\ba(t)\in\Delta\}=0$, the state $\rho(t+\Delta t|\ba(t)\in\Delta)$ is
indefinite, and let $\rho(t+\Delta t|\ba(t)\in\Delta)$ be an arbitrarily chosen 
density operator for mathematical convenience.

Suppose that the apparatus $\bA(\ba)$ measures a discrete 
observable $A$ with eigenvalues $a_{1},a_{2},\ldots$ and 
that the $\bA(\ba)$-measurement at the time $t$ is followed immediately 
by an $\bA(\bb)$-measurement measuring $B$.
The conditional probability of $\bb(t+\De t)\in\De'$ conditional
upon the outcome $\ba(t)=a_{n}$ is the probability of obtaining 
the outcome $\bb(t+\De t)\in\De'$ in the state $\rh(t+\De|\ba(t)=a_{n})$,
so that we have
\beqa\label{eq:0328e}
\lefteqn{\Pr\{\bb(t+\De t)\in\De'|\ba(t)=a_{n}\}}\quad\nn\\
&=&\Tr[E^{B}(\De')\rh(t+\De t|\ba(t)=a_{n})].
\eeqa
The joint probability distribution of the outcomes of the
successive measurement using $\bA(\ba)$ and $\bA(\bb)$ is
given by the well-known relation
\beqa\label{eq:0408b}
\lefteqn{\Pr\{\ba(t)\in\De,\bb(t+\De t)\in\De'\}}\nn\\
&=&
\sum_{a_{n}\in\De}\Pr\{\bb(t\!+\!\De t)\!\in\!\De'|\ba(t)\!=\!a_{n}\}
                  \Pr\{\ba(t)\!=\!a_{n}\}.
\eeqa

Now, suppose that $A$ is nondegenerate and that 
the apparatus $\bA(\ba)$ satisfies the repeatability hypothesis.
Then, the state of the system just after the 
$\bA(\ba)$-measurement conditional
upon the outcome $\ba(t)=a_{n}$ is determined uniquely as 
the normalized eigenstate
\beql{0408a}
\rh(t+\De t|\ba(t)=a_{n})=
\ket{\ph_{n}}\bra{\ph_{n}}
\eeq
corresponding to the eigenstate $a_{n}$, 
provided $\Pr\{\ba(t)=a_{n}\}>0$ \cite[pp.~215--217]{vN55}.

From \eq{0328e} and \eq{0408a}, we have
\beqas
\lefteqn{
\sum_{a_{n}\in\De}\Pr\{\bb(t+\De t)\in\De'|\ba(t)=a_{n}\}
\Pr\{\ba(t)=a_{n}\}}\quad\nn\\
&=&
\sum_{a_{n}\in\De}\bracket{\ph_{n}|E^{B}(\De')|\ph_{n}}
                  \bracket{\ph_{n}|\rh(t)     |\ph_{n}}\nn\\
&=&
\sum_{a_{n}\in\De}
\Tr[\ket{\ph_{n}}\bra{\ph_{n}}E^{B}(\De')
    \ket{\ph_{n}}\bra{\ph_{n}}\rh(t)].
\eeqas
Thus, from \eq{0408b} we have
\beqa\label{eq:0408c}
\lefteqn{\Pr\{\ba(t)\in\De,\bb(t+\De t)\in\De'\}}\nn\\
&=&
\sum_{a_{n}\in\De}
\Tr[\ket{\ph_{n}}\bra{\ph_{n}}E^{B}(\De')
    \ket{\ph_{n}}\bra{\ph_{n}}\rh(t)].
\eeqa
If $A$ and $B$ commute, we have
\beqas
\lefteqn{\Pr\{\ba(t)\in\De,\bb(t+\De t)\in\De'\}}\nn\\
&=&
\sum_{a_{n}\in\De}
\Tr[\ket{\ph_{n}}\bra{\ph_{n}}E^{B}(\De')\rh(t)]\\
&=&
\Tr[E^{A}(\De)E^{B}(\De')\rh(t)],
\eeqas
and hence we obtain the joint probability formula
\eq{0328c}.

By the above, we conclude that {\em if the apparatus $\bA(\ba)$ 
measuring a nondegenerate discrete observable $A$
satisfies the repeatability hypothesis,
the successive measurement using $\bA(\ba)$ and the apparatus 
$\bA(\bb)$ measuring an arbitrary observable $B$ commuting with
$A$ is a simultaneous measurement of $A$ and $B$.}

\subsection{Repeatability hypothesis and measuring processes}

Consider a model of measuring process in which an observable 
$A=\sum_{n}a_{n}\ket{\ph_{n}}\bra{\ph_{n}}$ with nondegenerate 
eigenvalues $a_{n}$ is measured by 
an apparatus $\bA(\ba)$ having the nondegenerate probe 
observable $B=\sum_{n}a_{n}\ket{\xi_{n}}\bra{\xi_{n}}$.
The {\em probe observable} is generally defined to be 
the quantum mechanical observable in the apparatus 
that is to be correlated with the measured observable 
by the measuring interaction 
and amplified in the later stage of the apparatus 
to the directly sensible variable read out eventually 
by the observer \cite{98PT}.  
In the conventional approach, the notion of the ``pointer position''
is used ambiguously instead, which sometimes means the ``probe observable'' 
and sometimes means the ``directly sensible variable''.
Let $\bP$ be the subsystem of the apparatus $\bA(\ba)$ that includes
the probe observable and actually interacts with the measured system $\bS$.
Suppose that $\bP$ is prepared in the state $\ket{\xi}$ just before
measurement.
Let $U$ be the unitary operator representing the time evolution
of the composite system $\bS+\bP$ during the measuring interaction.

The apparatus $\bA(\ba)$ 
satisfies the repeatability hypothesis if and only if $U$ satisfies 
\beql{1}
U:\ket{\ph_{n}}\otimes\ket{\xi}
\mapsto e^{i\th_{n}}\ket{\ph_{n}}\otimes\ket{\xi_{n}},
\eeq
where $e^{i\th_{n}}$ is an arbitrary phase factor.

\subsection{Measurements violating the repeatability hypothesis}

A typical model which does not satisfy \eq{1} 
is the photon counting measurement such that if the measurement is
taken place in the number state $\ket{\ph_{n}}=\ket{n}$ then the apparatus
absorbs all the photons and outputs the amplified classical 
energy proportional to the number of the absorbed photons.
In this case, $U$ satisfies 
\beql{2}
U:\ket{n}\otimes\ket{\xi}\mapsto\ket{0}\otimes\ket{\xi_{n}},
\eeq
and hence $U$ does not satisfy \eq{1}.

For less idealized models of photon counting measurement, 
we refer to \cite{SD81,IUO90}.
For the case of measurements of continuous observables,
there have been known also models of exact position measurement
that do not satisfy the repeatability hypothesis even approximately 
\cite{89RS,88MS,90QP}.  
They have been applied to the position 
monitoring that breaks the standard quantum limit \cite{88MS,Mad88}.

\subsection{Significance of the repeatability hypothesis}

Given $A$, $B$, and $\ket{\xi}$ generally, 
the apparatus $\bA(\ba)$ measures the 
observable $A$, or equivalently satisfies \eq{0328b}, 
if and only if $U$ satisfies
\beql{2.5}
U:\ket{\ph_{n}}\otimes\ket{\xi}\mapsto\ket{\ph'_{n}}\otimes\ket{\xi_{n}},
\eeq
where $\{\ket{\ph'_{n}}\}$ is an arbitrary family of normalized vectors,
not necessarily orthogonal.
Thus, if we do not assume the repeatability hypothesis,
the measurement is to correlate causally the input state $\ket{\ph_{n}}$
of the object before measurement with the output state $\ket{\xi_{n}}$ of 
the probe after measurement for some orthonormal basis $\{\ket{\xi_{n}}\}$.
On the other hand, the repeatability 
hypothesis requires not only that the input state $\ket{\ph_{n}}$ is 
correlated to the output state $\ket{\xi_{n}}$ causally 
but also that in the composite system after the measurement 
the input state $\ket{\ph_{n}}$ and the output state $\ket{\xi_{n}}$ 
is entangled to have the complete statistical correlation.

It is said quite often that the measurement of an observable $A$ is to
change the input state $\ket{\ps}$ to the eigenstate 
$\ket{\ph_{n}}$ with the probability 
$|\bracket{\ph_{n}|\ps}|^{2}$.
This does not follow from the Born statistical formula \eq{0328b}
but assumes the repeatability hypothesis. 
Thus, only when the measurement is assumed to satisfy the repeatability
hypothesis, we can say that the measurement changes probabilistically
the state of the object to one of the eigenstates of the measured 
observable.

Unless the repeatability hypothesis is assumed, it is, therefore, 
not a correct description that the measurement is 
to make a one-to-one correspondence (or to make an entanglement,
in the modern language)
between the state of the object before the measurement 
and the state of the probe after the measurement as described by \eq{1}, 
by which the problem of measuring the object is transferred to the problem
of measuring the probe \cite{LB39,Wig63}.

In the sequel, a measurement is called {\em repeatable} if it is 
carried out by the apparatus satisfying the repeatability hypothesis.

\section{Simultaneous measurements under the projection postulate}

\subsection{Von Neumann's measurements of degenerate observables}

For the observables with nondegenerate purely discrete spectrum,
the repeatability hypothesis determines the state after measurement
uniquely, but when the observable has degenerate eigenvalues, 
the state after measurement is not determined uniquely
but depends on the ``actual measuring arrangement'' \cite[p.~348]{vN55}.
Von Neumann \cite[p.~348]{vN55} considered the following ways of measurement
satisfying the repeatability hypothesis:
Let $\{\ket{\ph_{n,m}}\}$ be an orthonormal basis and let $A$ be an
observable represented by
\beql{3}
A=\sum_{n,m}a_{n}\ket{\ph_{n,m}}\bra{\ph_{n,m}}.
\eeq
Suppose that the observer performs a repeatable measurement of
a nondegenerate observable $A'$ given by 
\beql{4}
A'=\sum_{n,m}a_{n,m}\ket{\ph_{n,m}}\bra{\ph_{n,m}},
\eeq
where all $a_{n,m}$ are different, and that if the outcome of
the $A'$-measurement is $a_{n,m}$ then 
the outcome of the $A$-measurement is taken to be $a_{n}$.
Then we have a repeatable measurement of $A$. 

Suppose that the observable $A$ is measured at the time $t$
by the above way in a state (vector) $\ket{\ps}$ using the apparatus
$\bA(\ba)$, then at the time, $t+\De t$, just after the measurement
the object is left in the state (density operator)
\beqa\label{eq:5}
\lefteqn{\rh(t+\De t|\ba(t)=a_{n})}\quad\nn\\
&=&
\frac{1}{\sum_{m}|c_{n,m}|^{2}}
\sum_{m}|c_{n,m}|^{2}\ket{\ph_{n,m}}\bra{\ph_{n,m}},
\eeqa
where $c_{n,m}=\bracket{\ph_{n,m}|\ps}$.
This state depends not only on the observable $A$ and the 
outcome $a_{n}$ but also on the choice of the orthonormal basis
$\{\ket{\ph_{n,m}}\}$ that satisfies \eq{3}.
Since there are infinitely many essentially different choices of 
$\{\ket{\ph_{n,m}}\}$, the state change depends on the way of measurement
even if the repeatability hypothesis holds.

In this case, the joint probability distribution of the outcomes of
the $\bA(\ba)$-measurement and the immediately following 
$B$-measurement using the apparatus $\bA(\bb)$ is given by
\beqa
\lefteqn{\Pr\{\ba(t)\in\De,\bb(t+\De t)\in\De'\}}\nn\\
&=&
\!\sum_{a_{n}\in\De}\!\sum_{m}\!
\Tr[\ket{\ph_{n,m}}\bra{\ph_{n,m}}E^{B}(\De')
    \ket{\ph_{n,m}}\bra{\ph_{n,m}}\rh(t)].\nn\\
\eeqa
From the relation
$$
E^{A}(\De)=\sum_{a_{n}\in\De}\sum_{m}\ket{\ph_{n,m}}\bra{\ph_{n,m}},
$$
the joint probability formula \eq{0328c} holds for the arbitrary
input state $\rh(t)$ if and only if $A'$ and $B$ commute.
Although $A$ and $B$ commute, 
there are many choices of $\{\ket{\ph_{n,m}}\}$ such that $A'$ and $B$
do not commute.  Thus, even if $A$ and $B$ commute and $\bA(\ba)$
satisfies the repeatability hypothesis, the successive measurement using
$\bA(\ba)$ and $\bA(\bb)$ cannot be a simultaneous measurement 
of $A$ and $B$ in general.

\subsection{L\"{u}ders's formulation}

The previous argument shows the existence of infinitely many different
ways of measuring the same observable which satisfies the repeatability
hypothesis but does not satisfy the joint probability formula for
the simultaneous measurement.
Moreover, L\"{u}ders \cite{Lud51} has pointed out that the observable 
corresponding 
to the identity operator $I$ is considered to be measured without
changing the input state, but any one of the above measurement for
the identity changes the state unreasonably,
and he has suggested that the above measurement for a degenerate
observable is always more disturbing than the desirable one.

In order to determine the canonical way of measuring even the degenerate
observables, L\"{u}ders proposed the following hypothesis.

(P) {\em If an observable $A$ is measured in a state $\rh$, then
at the time just after measurement the object is left in
the state 
$$
\frac{E^{A}\{a\}\rh E^{A}\{a\}}{\Tr[E^{A}\{a\}\rh]},
$$
provided that the object leads to the outcome $a$ with 
$\Tr[E^{A}\{a\}\rh]>0$.}

In particular, if the object is measured in the vector state $\ket{\ps}$,
then the state after measurement is the vector state $E^{A}\{a\}\ket{\ps}$
up to normalization.  Thus, the eigenstate corresponding to the
outcome $a$ is uniquely chosen as the projection, and hence
the above hypothesis is called the {\em projection postulate}.

\subsection{Projection postulate and joint probability}

Suppose that a discrete observable $A$ of $\bS$ 
with eigenvalues $a_{1},a_{2},\ldots$
is measured at the time $t$ by the apparatus $\bA(\ba)$
measuring $A$ satisfying the projection postulate.
Then, the state of the system at the time $t+\De t$
just after the 
$\bA(\ba)$-measurement conditional
upon the outcome $\ba(t)=a_{n}$ is
\beql{0329d}
\rh(t+\De t|\ba(t)=a_{n})=
\frac{E^{A}\{a_{n}\}\rh(t) E^{A}\{a_{n}\}}
{\Tr[E^{A}\{a_{n}\}\rh(t)]},
\eeq
provided $\Pr\{\ba(t)=a_{n}\}>0$.

From \eq{0408b} and \eq{0329d}, we have
\beqa
\lefteqn{\Pr\{\ba(t)\in\De,\bb(t+\De t)\in\De'\}}\quad\nn\\
&=&
\sum_{a_{n}\in\De}\Tr[E^{B}(\De')E^{A}\{a_{n}\}\rh(t)E^{A}\{a_{n}\}].
\label{eq:0402a}
\eeqa
Thus, the joint probability formula \eq{0328c} holds for the
arbitrary input state $\rh(t)$ if and only if $A$ and $B$ commutes.
Therefore, we have seen the following theorem \cite{Lud51}.

{\bf Theorem 2.}
{\em The successive measurement of commuting observables $A$ and $B$ 
using apparatuses $\bA(\ba)$ and $\bA(\bb)$ is a simultaneous 
measurement of $A$ and $B$, if $\bA(\ba)$ satisfies the projection 
postulate.
}

We have seen that in order for a successive measurement of $A$ and $B$
to be a simultaneous measurement, the projection postulate
for the apparatus measuring $A$ is a sufficient condition.  
If we restrict our attention to the measuring apparatuses 
satisfying the projection postulate, any successive measurement of 
commuting observables is a simultaneous measurement,
but we have also seen that there are many kinds of measuring
apparatuses which do not satisfy the projection postulate.
Now, we shall turn to the problem as to under what condition 
a successive measurement of two or more observables 
can be considered as a simultaneous measurement.

\section{Nonselective operations of measuring apparatuses}

Every measuring process is considered to include an interaction,
called the {\em measuring interaction}, between the measured system 
and the measuring apparatus.
Let us consider the following description of the measuring interaction
arising in the measurement using the apparatus $\bA(\ba)$.  
In the following, the {\em probe} $\bP$ is a microscopic 
subsystem of the apparatus $\bA(\ba)$ that actually interacts with $\bS$.
More precisely, we define the probe $\bP$ to be the smallest 
subsystem of $\bA(\ba)$ such that the composite system $\bS+\bP$ is
isolated during the measuring interaction.
Since we assume naturally that $\bS+\bA(\ba)$ is isolated during the 
measuring interaction, the smallest subsystem exists.
The measurement is carried out by the interaction between the system 
$\bS$ and the probe $\bP$ and by the subsequent 
measurement on the probe $\bP$.
We assume that the probe system is a quantum mechanical system described
by the Hilbert space $\cK$ of state vectors.
At the time of measurement, $t$, the probe $\bP$ is in
the fixed state $\si$, so that the composite system is in the state
\beql{0328f}
\rh_{\bS+\bP}(t)=\rh(t)\otimes\si.
\eeq

The time evolution of the composite system
$\bS+\bP$ during the interaction is described by a unitary operator
$U$ on $\cH\otimes\cK$.  Hence, at the time just after the 
interaction, $t+\De t$, the composite system $\bS+\bP$ is 
in the state 
\beql{0328g}
\rh_{\bS+\bP}(t+\De t)=U(\rh(t)\otimes\si)U^{\da}.
\eeq
The outcome of this measurement is obtained by the measurement of
an observable $M$ of $\bP$ at the time $t+\De t$.
The observable $M$ is called the {\em probe observable}. 
Thus, the output probability distribution of the apparatus $\bA(\ba)$
is
\beql{0323a}
\Pr\{\ba(t)\in\De\}
=\Tr[(I\otimes E^{M}(\De))U(\rh(t)\otimes\si)U^{\da}].
\eeq

We shall call the above description of the measuring process as the
{\em indirect measurement model} determined by $(\cK,\si,U,M)$.
In this model, from \eq{0328g} the system ${\bf S}$ is in the state
\begin{equation}\label{eq:803a}
\rho(t+\Delta t)={\rm Tr}_{\cK}[U(\rho(t)\otimes\sigma)U^{\dagger}]
\end{equation}
at the time $t+\Delta t$, where ${\rm Tr}_{\cK}$ is the partial trace
over ${\cal K}$.
The state change $\rho(t)\mapsto\rho(t+\Delta t)$ is determined 
independent of the outcome of measurement and called
the {\em nonselective state change}.

As introduced previously, we have another type of state change
$\rh(t)\mapsto\rh(t+\De t|\ba(t)\in\De)$, where 
$\rho(t+\Delta t|\ba(t)\in\Delta)$ is the state at $t+\Delta t$ 
of ${\bf S}$ conditional upon $\ba(t)\in\Delta$.
Since the condition $\ba(t)\in\R$ makes no selection,
we have
\begin{equation}\label{eq:108b}
\rho(t+\Delta t|\ba(t)\in\R)=\rho(t+\Delta t).
\end{equation}
For $\Delta\not=\R$,
the state change $\rho(t)\mapsto\rho(t+\Delta t|\ba(t)\in\Delta)$ is called
the {\em selective state change}.

We define the transformation $\rh\mapsto \bT\rh$ 
for any trace class operator $\rh$ on the state space $\cH$ of
$\bS$ by the relation
\beql{803b}
\bT\rh=\Tr_{\cK}[U(\rh\otimes\si)U^{\da}].
\eeq
Then, $\bT$ is a trace preserving completely positive linear transformation 
on the space $\tc(\cH)$ of trace class operators on $\cH$
\cite{Dav76,Kra83,00MN}.
The transformation $\bT$ is determined by the apparatus preparation $\sigma$ 
and by the measuring interaction $U$, and is called the 
{\em nonselective operation}
of the apparatus ${\bf A}(\ba)$.
The nonselective operation $\bT$ represents the open system dynamics of the 
system $\bS$ from $t$ to $t+\De t$ and we have
\beql{803c}
\bT\rh(t)=\rh(t+\De t).
\eeq

As the converse of the definition \eq{803b}, 
it is well known that for every trace preserving
completely positive linear transformation on $\tc(\cH)$ 
there is an indirect measurement model such that $\bT$ is the nonselective
operation of that model \cite{83CR,84QC,Kra83}.

For any bounded linear transformation $\bL$ on $\tc(\cH)$,
its dual $\bL^{*}$ is defined by 
$$
\Tr[(\bL^{*}X)\rh]=\Tr[X(\bL\rh)]
$$
for all $\rh\in\tc(\cH)$ and $X\in\cL(\cH)$, where $\cL(\cH)$ stands
for the space of bounded operators on $\cH$.
Let $\bT^{*}$ be the dual of the nonselective operation $\bT$.
Then, $\bT^{*}$ is the normal unit preserving 
completely positive linear transformation on $\cL(\cH)$ such that 
\beql{971228a}
\Tr[(\bT^{*}X)\rh]=\Tr[X(\bT\rh)]
\eeq
for all $\rh\in\tc(\cH)$ and $X\in\cL(\cH)$ \cite[p.~18]{Dav76}.
We call $\bT^{*}$ the {\em dual nonselective operation}.
Let $X\in\cL(\cH)$ and $\rh\in\tc(\cH)$.
From \eq{803b} and a property of the partial trace, we have
\beqas
\Tr[X\bT\rh]
&=&\Tr[X\Tr_{\cK}[U(\rh\otimes\si)U^{\da}]\\
&=&\Tr[(X\otimes I)U(\rh\otimes\si)U^{\da}]\\
&=&\Tr[U^{\da}(X\otimes I)U(I\otimes\si)(\rh\otimes I)]\\
&=&\Tr[\Tr_{\cK}[U^{\da}(X\otimes I)U(I\otimes\si)]\rh].
\eeqas
Hence, from \eq{971228a} we have 
$$
\Tr[(\bT^{*}X)\rh]
=\Tr[\Tr_{\cK}[U^{\da}(X\otimes I)U(I\otimes\si)]\rh].
$$
Since $\rh$ is arbitrary, we have
\beql{0410a}
\bT^{*}X
=\Tr_{\cK}[U^{\da}(X\otimes I)U(I\otimes\si)]
\eeq
for all $X\in\cL(\cH)$.  
This characterizes the dual nonselective operation.

\section{Operation valued measures}

\subsection{Davies-Lewis postulates}

Davies and Lewis \cite{DL70} postulated that 
given an apparatus $\bA(\ba)$ for $\bS$, 
there is a mapping $\De\mapsto\bX(\De)$ 
from the Borel sets to the positive linear transformations
on $\tc(\cH)$
satisfying the following conditions:

(DL1) For any disjoint sequence of Borel sets $\Delta_{n}$ 
and for any $\rh\in\tc(\cH)$, 
\begin{mathletters}
\begin{equation}\label{eq:1230a}
\bX(\bigcup_{n}\De_{n})\rh=\sum_{n}\bX(\De_{n})\rh.
\end{equation}

(DL2) For any $\rho\in{\tau c}({\cal H})$,
\begin{equation}\label{eq:1230b}
{\rm Tr}[\bX(\R)\rh]={\rm Tr}[\rho].
\end{equation}

(DL3) For any Borel set $\Delta$,
\begin{equation}\label{eq:1230c}
\Pr\{\ba(t)\in\De\}={\rm Tr}[\bX(\De)\rh(t)].
\end{equation}

(DL4) For any Borel set $\Delta$ with $\Pr\{\ba(t)\in\Delta\}>0$,
\begin{equation}\label{eq:1230d}
\rho(t+\Delta t|\ba(t)\in\Delta)
=\frac{\bX(\De)\rho(t)}{{\rm Tr}[\bX(\De)\rho(t)]}.
\end{equation}
\end{mathletters}

We call the above mapping $\bX:\De\mapsto\bX(\De)$ the {\em operation
valued measure} or the {\em operational distribution} of the
apparatus $\bA(\ba)$.
In general, we call any bounded linear transformation on $\tc(\cH)$ 
a {\em superoperator} for $\cH$.  
Any mapping $\De\mapsto\bX(\De)$
from the Borel sets to the positive superoperators for $\cH$
is called a {\em positive superoperator valued (PSV) measure}
if it satisfies condition (DL1).
Moreover, it is called {\em normalized}
if it satisfies condition (DL2).
Accordingly, the operation valued measure of $\bA(\ba)$
is the normalized PSV measure satisfying (DL3) and (DL4).

The validity of the Davies-Lewis postulate for the apparatuses with
indirect measurement models was previously demonstrated {\em based on} 
the joint probability formula in \cite{83CR,84QC}, where it is also 
shown that any normalized PSV measures which are realizable by indirect 
measurement models are completely positive and {\em vice versa}.

\subsection{Determination of operation valued measures based on
the projection postulate}

In order to determine the operation valued measure corresponding
to the given measuring apparatus,
we need to describe the measuring process by an indirect measurement model.
Then the measurement is divided into the two processes,
the measuring interaction in the object-probe composite system
and the probe measurement.
Given the indirect measurement model, the current formulation has 
two arguments to determine the operation valued measure:
one relies on the projection postulate (cf.~\cite{Kra83} for
yes-no measurements) and the other relies on the joint probability 
formula \cite{83CR,84QC,89RS}.

In the first approach, the probe measurement 
is assumed explicitly to satisfy the projection postulate.
Consequently, the operation valued measure is determined by the 
unitary operator of the measuring interaction and the projection
operator derived by the projection postulate with partial trace
over the probe.  

The argument runs as follows. 
Assume that the apparatus $\bA(\ba)$ has the indirect measurement 
model $(\cK,\si,U,M)$, where the probe observable $M$ is purely 
discrete with eigenvalues $a_{n}$.
Let us suppose that the measuring interaction between the system
$\bS$ and the apparatus $\bA(\ba)$ is turned on from $t$ to $t+\De t$ 
and that the observer measures the probe observable $M$ at the 
time $t+\De t$ using the apparatus $\bA(\bm)$.
Let $t+\De t+\ta$ be the time just after the measuring interaction
is turned off
between the probe $\bP$ and the apparatus $\bA(\bm)$.
The system $\bA(\bm)$ is considered as a subsystem of $\bA(\ba)$
including the later stages after the probe $\bP$.
Assume that the apparatus $\bA(\bm)$ satisfies the projection postulate
and that the outcome is $\bm(t+\De t)=a_{n}$.
Then, at the time $t+\De t+\ta$ the composite system $\bS+\bP$ is
in the state 
\beqa\label{eq:0411a}
\lefteqn{\rh_{\bS+\bP}(t+\De t+\ta|\bm(t+\De t)=a_{n})}\quad\nn\\
&=&\frac{(I\otimes E^{M}\{a_{n}\})\rh_{\bS+\bP}(t+\De t)
         (I\otimes E^{M}\{a_{n}\})}
    {\Tr[(I\otimes E^{M}\{a_{n}\})\rh_{\bS+\bP}(t+\De t)]}.
\eeqa 
Since the outcome of the $\bA(\bm)$-measurement at the time $t+\De t$
is interpreted as the outcome of the $\bA(\ba)$-measurement at the
time $t$, the condition $\bm(t+\De t)=a_{n}$ is equivalent to the
condition $\ba(t)=a_{n}$.
It follows that at the time $t+\De t+\ta$ the system $\bS$ is in the state
\beqa\label{eq:0411b}
\lefteqn{\rh(t+\De t+\ta|\ba(t)=a_{n})}\quad\nn\\
&=&
\Tr_{\cK}[\rh_{\bS+\bP}(t+\De t+\ta|\bm(t+\De t)=a_{n})].
\eeqa
From \eq{0328g}, \eq{0411a}, and \eq{0411b}, we have
\beqa\label{eq:0411d'}
\lefteqn{\rh(t+\De t+\ta|\ba(t)=a_{n})}\nn\\
&=&\frac{\Tr_{\cK}[(I\otimes E^{M}\{a_{n}\})U(\rh(t)\otimes\si)U^{\da}
                   (I\otimes E^{M}\{a_{n}\})]}
              {\Tr[(I\otimes E^{M}\{a_{n}\})U(\rh(t)\otimes\si)U^{\da}]}.
\nn\\
\eeqa
By the well-known relation
\beql{p-tr}
\Tr_{\cK}[(I\otimes X)Y]=\Tr_{\cK}[Y(I\otimes X)]
\eeq
for all $X\in\cL(\cK)$ and $Y\in\cL(\cH\otimes\cK)$, we have
\beqa
\lefteqn{\Tr_{\cK}[(I\otimes E^{M}\{a_{n}\})U(\rh(t)\otimes\si)U^{\da}
                   (I\otimes E^{M}\{a_{n}\})]}\qquad\qquad\nn\\
&=&      \Tr_{\cK}[(I\otimes E^{M}\{a_{n}\})U(\rh(t)\otimes\si)U^{\da}].
\eeqa
Hence, we have an important relation
\beqa\label{eq:0411d}
\lefteqn{\rh(t+\De t+\ta|\ba(t)=a_{n})}\quad\nn\\
&=&\frac{\Tr_{\cK}[(I\otimes E^{M}\{a_{n}\})U(\rh(t)\otimes\si)U^{\da}]}
              {\Tr[(I\otimes E^{M}\{a_{n}\})U(\rh(t)\otimes\si)U^{\da}]}.
\eeqa
Let $\De$ be an arbitrary Borel set such that $\Pr\{\ba(t)\in\De\}>0$.
Then, we have naturally
\beqa\label{eq:0411e}
\lefteqn{\rh(t+\De t+\ta|\ba(t)\in\De)}\nn\\
&=&
\frac{\sum_{a_{n}\in\De}\Pr\{\ba(t)=a_{n}\}\rh(t+\De t+\ta|\ba(t)=a_{n})}
     {\Pr\{\ba(t)\in\De)\}}.\nn\\
\eeqa
From \eq{0323a}, \eq{0411d}, and \eq{0411e}, we have
\beqa\label{0411f}
\lefteqn{\rh(t+\De t+\ta|\ba(t)\in\De)}\nn\quad\\
&=&
\frac
{\Tr_{\cK}[(I\otimes E^{M}(\De))U(\rh(t)\otimes\si)U^{\da}]}
      {\Tr[(I\otimes E^{M}(\De))U(\rh(t)\otimes\si)U^{\da}]}.
\eeqa
To obtain the final result,
suppose that the $\bA(\bm)$-measurement is instantaneous, i.e., 
$\ta\approx 0$, and that there is no interaction between $\bS$ 
and the outside of $\bA(\ba)$ from $t$ to $t+\De t+\ta$.
Then, in this time interval the state changes of $\bS$ are negligible 
due to the $\bA(\bm)$-measurement, the time evolution of $\bS$,
and the decoherence from the environment.  Consequently, we have
\beql{0411c}
\rh(t+\De t|\ba(t)\in\De)
=
\rh(t+\De t+\ta|\ba(t)\in\De).
\eeq
Therefore, we have reached the final form
\beqa\label{eq:0412a}
\lefteqn{\rh(t+\De t|\ba(t)\in\De)}\nn\quad\\
&=&
\frac
{\Tr_{\cK}[(I\otimes E^{M}(\De))U(\rh(t)\otimes\si)U^{\da}]}
      {\Tr[(I\otimes E^{M}(\De))U(\rh(t)\otimes\si)U^{\da}]}.
\eeqa

From the above, the operation valued measure of $\bA(\ba)$ is 
determined by
\beql{0412b}
\bX(\De)\rh=
{\Tr_{\cK}[(I\otimes E^{M}(\De))U(\rh\otimes\si)U^{\da}]}
\eeq
for all Borel sets $\De$ and all trace class operators $\rh$.
Furthermore, it follows easily from properties of partial trace 
that $\bX$ satisfies conditions (DL1)--(DL4).

Obviously, the above approach explicitly excludes the possibility of 
measuring the probe by the apparatus not satisfying
the projection postulate such as a photon counter.
Accordingly, this approach cannot apply correctly to any 
measurements with continuous probe observables such as the 
position of the probe pointer, since no apparatuses measuring 
continuous observables satisfy the repeatability hypothesis 
\cite{84QC,85CA}. 
Moreover, the argument assumes that the probe measurement should
be instantaneous.

\subsection{Determination of operation valued measures based on
the joint probability formula}

In the second approach, generalizing von Neumann's argument on 
repeated measurements of the same observable \cite[pp.~211--223]{vN55}, 
it is assumed that the observer were to measure an arbitrary 
observable of the object system again at the time just after the measuring 
interaction and then considered is the joint probability distribution
of the outcomes of the probe measurement and the second object 
measurement \cite{83CR,84QC,89RS}.
By assuming that the above joint probability distribution satisfies 
the joint probability formula for the simultaneous measurement, 
we can determine the operation valued measure.

Since the joint probability formula is well formulated even in the case
where the probe observable has continuous spectrum, the second 
approach can be applied to measurements of continuous observables.
Moreover, in the case of the discrete probe observable, the second
approach leads to the same operation valued measure as 
the first approach, so that the second approach is consistent with 
the first.

The argument in the second approach runs as follows.
Let $\bA(\ba)$ be an apparatus described by the indirect measurement
model $(\cK,\si,U,M)$.
Suppose that the system $\bS$ is measured at the time $t$
by the apparatus $\bA(\ba)$.
Suppose that at the time $t+\Delta t$ just after the measuring
interaction, the observer were to measure an arbitrary observable $B$ 
of the same object ${\bf S}$ by an apparatus $\bA(\bb)$.
The conditional probability of $\bb(t+\De t)\in\De'$ given $\ba(t)\in\De$ 
is the probability of $\bb(t+\De t)\in\De'$ 
in the state $\rh(t+\De|\ba(t)\in\De)$,
so that the joint probability distribution of $\ba(t)$ and $\bb(t+\De t)$
satisfies
\begin{eqnarray}\label{eq:8}
\lefteqn{\Pr\{\ba(t)\in\Delta,\bb(t+\Delta t)\in\Delta'\}}\quad\nonumber\\
&=&{\rm Tr}[E^{B}(\Delta')\rho(t+\Delta t|\ba(t)\in\Delta)]
\Pr\{\ba(t)\in\De\}.
\end{eqnarray}
For any Borel set $\Delta$, 
let $\bX(\De,\rho(t))$ be the trace class operator defined by
\begin{equation}\label{eq:T2}
\bX(\De,\rho(t))
=\Pr\{\ba(t)\in\De\}\rho(t+\Delta t|\ba(t)\in\Delta).
\end{equation}
From (\ref{eq:8}) we have
\begin{eqnarray}\label{eq:T3}
\lefteqn{\Pr\{\ba(t)\in\Delta,\bb(t+\Delta t)\in\Delta'\}}\quad\nn\\
&=&{\rm Tr}[E^{B}(\Delta')\bX(\De,\rho(t))].
\end{eqnarray}

On the other hand, by the indirect measurement model the output
$\ba(t)$ of this measurement is obtained by the measurement of 
the probe observable $M$ at the time $t+\De t$.
Let $\bA(\bm)$ be the apparatus measuring $M$ at the time $t+\De t$.
Then, the probabilistic event ``$\ba(t)\in\De$'' 
is equivalent to the probabilistic event ``$\bm(t+\De t)\in\De$''
and hence we have
\beqa\label{0403a}
\lefteqn{\Pr\{\ba(t)\in\De,\bb(t+\De t)\in\De'\}}\quad\nn\\
&=&\Pr\{\bm(t+\De t)\in\De,\bb(t+\De t)\in\De'\}.
\eeqa
Since the observable $M$ of $\bP$ and the observable $B$ of $\bS$ 
are simultaneously measurable, 
if the $\bA(\bm)$-measurement and the $\bA(\bb)$-measurement
can be considered as a simultaneous measurement, we have
\beqas
\lefteqn{\Pr\{\bm(t+\De t)\in\De,\bb(t+\De t)\in\De'\}}\quad\\
&=&
\Tr[(E^{B}(\De')\otimes E^{M}(\De))\rh_{\bS+\bP}(t+\De t)]\nn\\
&=&
\Tr[(E^{B}(\De')\otimes E^{M}(\De))U(\rh(t)\otimes\si)U^{\da}].
\eeqas
By the property of partial trace, we have
\beqa\label{eq:0403b}
\lefteqn{\Pr\{\bm(t+\De t)\in\De,\bb(t+\De t)\in\De'\}}\quad\nn\\
&=&
\Tr[E^{B}(\De')\Tr_{\cK}[(I\otimes E^{M}(\De))U(\rh(t)\otimes\si)U^{\da}].
\eeqa
Since $B$ and $\De'$ are arbitrary, 
from \eq{T3}--\eq{0403b} we have
\beql{0403c}
\bX(\De,\rho(t))
=
\Tr_{\cK}[(I\otimes E^{M}(\De))U(\rh(t)\otimes\si)U^{\da}].
\eeq
Suppose that $\Pr\{\ba(t)\in\De\}>0$.
From \eq{T2}, we have 
\beqas
\lefteqn{\rh(t+\De t|\ba(t)\in\De)}\quad\\
&=&
\frac{\bX(\De,\rho(t))}{\Tr[\bX(\De,\rho(t))]}\\
&=&
\frac
{\Tr_{\cK}[(I\otimes E^{M}(\De))U(\rh(t)\otimes\si)U^{\da}]}
      {\Tr[(I\otimes E^{M}(\De))U(\rh(t)\otimes\si)U^{\da}]}.
\eeqas
Hence, we have shown that relation \eq{0412a} holds 
for the apparatus $\bA(\ba)$ given in this argument. 
Let $\bX$ be the mapping $\De\mapsto\bX(\De)$ defined by relation
\eq{0412b} for the present apparatus.  
Then, $\bX$ satisfies conditions
(DL1)--(DL2) by the properties of partial trace
as before.
From \eq{0403c} we have
\beq
\bX(\De)\rh(t)=\bX(\De,\rh(t))
\eeq
and hence $\bX$ satisfies conditions (DL3)--(DL4).
Thus, $\bX$ satisfies the Davies-Lewis postulate for 
the apparatus $\bA(\ba)$ given above.

We have shown that the determination \eq{0412b} of the operation
valued measure holds without assuming the projection postulate
for the probe measurement. 
Nevertheless, in order to justify the formula \eq{0412b} generally 
we need to justify the joint probability formula without assuming the
projection postulate.
This put a serious constraint on the theoretical device to explore
our problem.  Indeed, because of the threat of the circular argument,
the above arguments do not enable us to take advantage of the 
operation valued measures for the justification of the joint 
probability formula.
In the conventional measurement theory, the similar kind of 
circular argument has been known as the infinite regress of 
the von Neumann chain.
Despite the above difficulties, 
we shall show, in the following sections, an alternative approach
without any fear of the circular argument.

\section{Statistical approach to the operation valued measures} 

\subsection{Existence of the operation valued measures}

In what follows, we shall prove the Davies-Lewis 
postulate {\em without} assuming the joint probability formula
or the projection postulate.

Let us suppose that the system $\bS$ is measured at the time $t$ 
by the apparatus $\bA(\ba)$ and at the time $t+\De t$ immediately
after this measurement an observable $B$ of $\bS$ is measured 
using an apparatus $\bA(\bb)$.
Then, the joint probability distribution of the outcomes
of the $A$-measurement and the $B$-measurement satisfies \eq{8}.
For any Borel set $\Delta$, 
let $\bX(\Delta,\rho(t))$ be the trace class operator defined by
\eq{T2}.  
Then, from (\ref{eq:8}), $\bX(\De,\rh(t))$ satisfies \eq{T3}.
Since the input state $\rh(t)$ is assumed to be an arbitrary
density operator, (\ref{eq:T2}) defines the transformation 
$\bX(\Delta)$ that maps $\rho(t)$ to $\bX(\Delta,\rho(t))$.
From \eq{T2} and \eq{T3}, $\bX(\De)$ satisfies the relations
\beq
\bX(\De)\rh(t)
=\Pr\{\ba(t)\in\De\}\rh(t+\De t|\ba(t)\in\De)
\label{eq:T2'}
\eeq
and
\beq
\Pr\{\ba(t)\in\Delta,\bb(t+\Delta t)\in\Delta'\}
=\Tr[E^{B}(\Delta')\bX(\Delta)\rho(t)].
\label{eq:T3'}
\eeq

Suppose that the input state $\rho(t)$ is a mixture of 
density operators $\rho_{1}$ and
$\rho_{2}$, i.e., 
\begin{equation}\label{eq:T4}
\rho(t)=\alpha\rho_{1}+(1-\alpha)\rho_{2}
\end{equation}
where $0<\alpha<1$.
This means that at the time $t$ 
the measured object ${\bf S}$ is sampled 
randomly from an ensemble of similar systems described by the
density operator $\rho_{1}$ with probability $\alpha$ and from 
another ensemble described by the density operator $\rho_{2}$
with probability $1-\alpha$.
Thus we have naturally
\begin{eqnarray}\label{eq:T4.5}
\lefteqn{\Pr\{\ba(t)\in\Delta,\bb(t+\Delta t)\in\Delta'|\rho(t)
=\alpha\rho_{1}+(1-\alpha)\rho_{2}\}}\nonumber\\
&=&\alpha\Pr\{\ba(t)\in\Delta,\bb(t+\Delta t)\in\Delta'|\rho(t)
=\rho_{1}\}\nonumber\\
& &\mbox{ }+(1-\alpha)\Pr\{\ba(t)\in\Delta,\bb(t+\Delta t)\in\Delta'|\rho(t)
=\rho_{2}\},\nonumber\\
\end{eqnarray}
where $\Pr\{E|F\}$ stands for the conditional probability
of $E$ given $F$.   
From (\ref{eq:T3'}) and (\ref{eq:T4.5}), we have
\begin{eqnarray*}
\lefteqn{{\rm Tr}\left[E^{B}(\Delta')\bX(\De)
\left[\alpha\rho_{1}+(1-\alpha)\rho_{2}\right]\right]}\quad\\
&=&\alpha\Tr[E^{B}(\De')\bX(\De)\rho_{1}]\\
& &\mbox{ }+(1-\alpha)\Tr[E^{B}(\De')\bX(\De)\rho_{2}]\\
&=&{\rm Tr}
[E^{B}(\Delta')[\alpha \bX(\De)\rh_{1}+(1-\alpha)\bX(\De)\rh_{2}]].
\end{eqnarray*}
Since $B$ and $\Delta'$ are arbitrary, we have
\begin{equation}\label{eq:T6}
\bX(\De)\left[\alpha\rho_{1}+(1-\alpha)\rho_{2}\right]
=\alpha \bX(\De)\rh_{1}+(1-\alpha)\bX(\De)\rh_{2}.
\end{equation}
It follows that $\bX(\De)$ is an affine transformation 
from the space of density 
operators to the space of trace class operators, 
so that it can be extended to a unique positive 
superoperator \cite{00MN}.

We have proved that for any apparatus ${\bf A}(\ba)$ measuring $A$
there is uniquely a family $\{\bX(\De)|\ \Delta\in{\cal B}(\R)\}$
of positive superoperators
such that (\ref{eq:T2'}) and (\ref{eq:T3'}) hold,
where ${\cal B}(\R)$ stands for the collection of all Borel sets.

By the countable additivity of probability, 
if $\Delta=\bigcup_{n}\Delta_{n}$ for disjoint Borel sets $\Delta_{n}$, 
we have
\begin{eqnarray}
\lefteqn{
\Pr\{\ba(t)\in\Delta,\bb(t+\Delta t)\in\Delta'\}
}\qquad\nonumber\\
&=&
\sum_{n}\Pr\{\ba(t)\in\Delta_{n},\bb(t+\Delta t)\in\Delta'\}.
\label{eq:108a}
\end{eqnarray}
By (\ref{eq:T3'}) and (\ref{eq:108a}), we have
\beqas
\Tr[E^{B}(\De')\bX(\De)\rh(t)]
&=&
\sum_{n}\Tr[E^{B}(\De')\bX(\De_{n})\rh(t)]\\
&=&
\Tr[E^{B}(\De')\sum_{n}\bX(\De_{n})\rh(t)].
\eeqas
Since $B$ and $\De'$ are arbitrary, we have 
$$
\bX(\De)\rh(t)=\sum_{n}\bX(\De_{n})\rh(t).
$$
Since $\rh(t)$ is arbitrary, condition (DL1) holds for arbitrary 
density operator $\rh$ and hence by linearity, condition (DL1) holds
for all $\rh\in\tc(\cH)$.
Conditions (DL3) and (DL4) are obvious from \eq{T2}.
From (DL3), we have
$$
\Tr[\bX(\R)\rh(t)]=1.
$$
Since $\rh(t)$ is arbitrary, condition (DL2) holds for arbitrary 
density operator $\rh$ and hence by linearity, condition (DL2) holds
for all $\rh\in\tc(\cH)$.
Thus, the mapping $\bX:\De\mapsto\bX(\De)\rh$ 
satisfies the Davies-Lewis postulate.

It should be noted that the present derivation rely on
neither the existence of the indirect measurement model, 
the joint probability formula, nor the projection postulate.
The crucial assumption in the above argument is \eq{T4.5} which
follows from the basic principle underlying the notion of the mixture 
of states.
Thus, we can conclude that {\em every measuring apparatus has 
the operation valued measure satisfying the Davies-Lewis
postulate}.

\subsection{Basic properties of the operation valued measures}

Let $\bA(\ba)$ be a measuring apparatus for the system $\bS$
with the operation valued measure $\bX$.
Let us assume that the apparatus $\bA(\ba)$ measures an observable
$A$.  In this case, from \eq{0328b} and (DL3) we have
\beql{0412d}
\Tr[\bX(\De)\rh(t)]=\Tr[E^{A}(\De)\rh(t)].
\eeq
Let $\bX(\De)^{*}$ be the dual of $\bX(\De)$.  Then, we have
$$
\Tr[(\bX(\De)^{*}I)\rh(t)]=\Tr[E^{A}(\De)\rh(t)].
$$
Since $\rh(t)$ is arbitrary, we conclude
\beql{0412e}
\bX(\De)^{*}I=E^{A}(\De)
\eeq
for any Borel set $\De$.

We say that a PSV measure $\bX$ is 
{\em $A$-compatible} if $\bX$ satisfies relation \eq{0412e}.
By the above, the operation valued measure of the apparatus $\bA(\ba)$
measuring $A$ is an $A$-compatible PSV measure.

Now we are ready to state the following important relations
for operation valued measures.

{\bf Theorem 3.}
{\em Let $A$ be an observable and let $\bX$ be an $A$-compatible 
PSV measure.
Then, for any Borel set $\Delta$ and any trace class operator
$\rho$ we have
\begin{eqnarray}\label{eq:127b}
\bX(\De)\rho
&=&\bX(\R)\left(E^{A}(\Delta)\rho\right)
=\bX(\R)\left(\rho E^{A}(\Delta)\right)\nonumber\\
&=&\bX(\R)\left(E^{A}(\Delta)\rho E^{A}(\Delta)\right),
\end{eqnarray}
and for any bounded operator $B$ we have
\begin{eqnarray}\label{eq:127b'}
\bX(\De)^{*}B
&=&(\bX(\R)^{*}B)E^{A}(\De)
=E^{A}(\Delta)\bX(\R)^{*}B\nonumber\\
&=&E^{A}(\Delta)(\bX(\R)^{*}B)E^{A}(\Delta).
\end{eqnarray}
}

A proof of the above theorem was given in \cite{84QC} for the 
case where $\bX(\De)$ is completely positive, and another proof
was given in \cite{97OQ} for the case where $A$ is discrete.
The general proof necessary for the above theorem runs as 
follows \cite{00MO}.

{\em Proof.}
Let $C$ be a bounded operator such that $0\le C\le I$
and let $\De\in\cB(\R)$.
We define
$$
\begin{array}{ll}
A_{11}=\bX(\De)^{*}C,\quad&
A_{12}=\bX(\De)^{*}(I-C),\\
A_{21}=\bX(\R\setminus\De)^{*}C,\quad&
A_{22}=\bX(\R\setminus\De)^{*}(I-C),\\
P_{1}=E^{A}(\De),\quad&
P_{2}=I-E^{A}(\De),\\
Q_{1}=\bX(\R)^{*}C,\quad&                 
Q_{2}=I- \bX(\R)^{*}C.
\end{array}
$$
Then, for $i,j=1,2$ we have $0\le A_{ij}\le P_{i}$, 
so that $[A_{ij},P_{i}]=[A_{ij},P_{j}]=0$.
It follows that $Q_{j}=A_{1j}+A_{2j}$ commutes with $P_{1}$
and $P_{2}$ as well.
Thus, 
$$
A_{ij}=P_{i}A_{ij}\le P_{i}Q_{j}.
$$
On the other hand, we have $\sum_{ij}A_{ij}=I$ and 
$\sum_{ij}P_{i}Q_{j}=I$, whence $A_{ij}=P_{i}Q_{j}$.
It follows that 
$$
\bX(\De)^{*}C=E^{A}(\De)\bX(\R)^{*}C.
$$
By taking adjoint, we also have 
$$
\bX(\De)^{*}C=(\bX(\R)^{*}C)E^{A}(\De).
$$
Since any bounded operator $B$ can be represented by
$B=\sum_{n=0}^{3}\la_{n}C_{n}$ with positive operators 
$0\le C_{n}\le I$ and complex numbers $\la_{n}$, we have
$$
\bX(\De)^{*}B=E^{A}(\De)\bX(\R)^{*}B=(\bX(\R)^{*}B)E^{A}(\De)
$$
for any $\De\in\cB(\R)$ and $B\in\cL(\cH)$.
By multiplying $E^{A}(\De)$ from the both sides, we also have
$$
\bX(\De)^{*}B=E^{A}(\De)(\bX(\R)^{*}B)E^{A}(\De).
$$
Hence, relations \eq{127b'} hold.  Relations \eq{127b}
follow easily by taking the duals of $\bX(\De)^{*}$
and $\b\bX(\R)^{*}$.
\qed

By the above theorem, the operation valued measure
$\bX$ of an arbitrary apparatus $\AA$ measuring
$A$ is determined uniquely by 
the nonselective operation $\bT=\bX(\R)$ of $\AA$.

Mathematical theory of PSV measures
was introduced by Davies and Lewis \cite{DL70}
based on conditions (DL1) and (DL2) as mathematical axioms;
see also Davies \cite{Dav76}.
Their relations with measuring processes were established in
\cite{84QC,85CA,85CC,86IQ,83CR,95MM} and applied to analyzing
various measuring processes in \cite{89RS,88MS,90QP,93CA}.

\subsection{Operation valued measures of indirect measurement models}

Suppose that the apparatus $\bA(\ba)$ measuring $A$ has the indirect 
measurement model $(\cK,\si,U,M)$.  
In this case, we can determine the operation valued
measure $\bX$ of the apparatus $\bA(\ba)$ without assuming
the joint probability distribution or the projection postulate,
as follows.

Let $\bX$ be the operation valued measure of the apparatus $\bA(\ba)$.
Then, $\bX$ satisfies (DL1)--(DL4) and hence $\bX$ is an $A$-compatible
PSV measure. It follows from Theorem {3} that $\bX$ satisfies
\beql{0414a}
\bX(\De)\rh=\bX(\R)[E^{A}(\De)\rh],
\eeq
where $\De\in\cB(\R)$ and $\rh\in\tc(\cH)$.
Since $\bA(\ba)$ has the indirect measurement model $(\cK,\si,U,M)$,
relation \eq{803a} holds.
By (DL4), \eq{803a}, and \eq{108b}, we have 
$$
\bX(\R)\rh(t)=\Tr_{\cK}[U(\rh(t)\otimes\si)U^{\da}].
$$
Since $\rh(t)$ is arbitrary and $\bX(\R)$ is linear, 
the above relation can be extended to trace class operators $\rh$ 
from density operators $\rh(t)$, 
so that we have
\beql{0414b}
\bX(\R)\rh=\Tr_{\cK}[U(\rh\otimes\si)U^{\da}]
\eeq
for all $\rh\in\tc(\cH)$.  Now, we consider the expression
\beql{0414c}
\cE(\De)\rh=\Tr_{\cK}[(I\otimes E^{M}(\De))U(\rh\otimes\si)U^{\da}],
\eeq
where $\De\in\cB(\R)$ and $\rh\in\tc(\cH)$.
Then, we can show purely mathematically that the mapping $\cE:
\De\mapsto\cE(\De)$ defined above is an $A$-compatible PSV measure
satisfying
\beql{0414d}
\cE(\R)\rh=\Tr_{\cK}[U(\rh\otimes\si)U^{\da}].
\eeq
Thus, $\cE$ satisfies the assumptions of Theorem {3}, 
and hence we have
\beql{0414e}
\cE(\De)\rh=\cE(\R)[E^{A}(\De)\rh].
\eeq
From \eq{0414b} and \eq{0414d}, we have 
\beql{0414f}
\cE(\R)=\bX(\R)
\eeq
and hence from \eq{0414a} and \eq{0414e} we have
\beqas
\cE(\De)\rh
&=&\cE(\R)[E^{A}(\De)\rh]\\
&=&\bX(\R)[E^{A}(\De)\rh]\\
&=&\bX(\De)\rh.
\eeqas
Therefore, we conclude that $\bX$ satisfies \eq{0412b}.

From \eq{0412b} and \eq{127b}, we have the following
expressions for $\bX$:
\begin{mathletters}
\beqa
\bX(\De)\rh
&=&\Tr_{\cK}[(I\otimes E^{M}(\De))U(\rh\otimes\si)U^{\da}]\\
&=&{\rm Tr}_{{\cK}}[U(\rho(t)E^{A}(\Delta)\otimes\sigma)U^{\dagger}]\\
&=&{\rm Tr}_{{\cK}}[U(E^{A}(\Delta)\rho(t)\otimes\sigma)U^{\dagger}]\\
&=&{\rm Tr}_{{\cK}}[U(E^{A}(\Delta)\rho(t)E^{A}(\Delta)
\otimes\sigma)U^{\dagger}].
\eeqa
\end{mathletters}
Thus, if $\Pr\{\ba(t)\in\Delta\}>0$, we obtain the following relations.
\begin{mathletters}
\label{eq:108d}
\begin{eqnarray}
\lefteqn{\rho(t+\Delta t|\ba(t)\in\Delta)}\qquad\nonumber\\
&=&\frac{{\rm Tr}_{{\cK}}[(I\otimes E^{M}(\De))
U(\rho(t)\otimes\sigma)U^{\dagger}]}
{{\rm Tr}[E^{A}(\Delta)\rho(t)]}\label{eq:108d-a'}\\
&=&\frac{{\rm Tr}_{{\cK}}[U(\rho(t)E^{A}(\Delta)\otimes\sigma)U^{\dagger}]}
{{\rm Tr}[E^{A}(\Delta)\rho(t)]}\label{eq:108d-a}\\
&=&\frac{{\rm Tr}_{{\cK}}[U(E^{A}(\Delta)\rho(t)\otimes\sigma)U^{\dagger}]}
{{\rm Tr}[E^{A}(\Delta)\rho(t)]}\label{eq:108d-b}\\
&=&\frac{{\rm Tr}_{{\cK}}[U(E^{A}(\Delta)\rho(t)E^{A}(\Delta)
\otimes\sigma)U^{\dagger}]}
{{\rm Tr}[E^{A}(\Delta)\rho(t)]}.\label{eq:108d-c}
\end{eqnarray}
\end{mathletters}

\section{Disturbance in measurement}

\subsection{Disturbance and simultaneous measurability}

Let $B$ be an arbitrary observable of ${\bf S}$.
We say that the measurement using an apparatus $\AA$ {\em does not disturb} 
the observable $B$ if the nonselective
state change does not perturb the probability distribution of $B$,
that is, we have
\begin{equation}\label{eq:1224a}\label{eq:9-1}
{\rm Tr}[E^{B}(\Delta)\rho(t+\Delta t)]
={\rm Tr}[E^{B}(\Delta)e^{-iH\Delta t/\hbar}\rho(t)e^{iH\Delta t/\hbar}]
\end{equation}
for any Borel set $\Delta$, where $H$ is the Hamiltonian 
of the system ${\bf S}$.
The measurement is said to be {\em instantaneous} if the duration 
$\Delta t$ of the measurement is negligible in the time scale 
of the time evolution of the system $\bS$.
Thus, the instantaneous measurement using the apparatus $\bA(\ba)$ 
does not disturb $B$ if and only if
\begin{equation}\label{eq:10}\label{eq:9-2}
{\rm Tr}[E^{B}(\Delta)\rho(t+\Delta t)]={\rm Tr}[E^{B}(\Delta)\rho(t)]
\end{equation}
for any Borel set $\Delta$.

Let $\bX$ be the operation valued measure of the apparatus $\bA(\ba)$
and $\bT=\bX(\R)$ be the nonselective operation of $\bA(\ba)$.
Then, from (DL4) we have
\beql{9-3}
\rh(t+\De t)=\bT\rh(t)
\eeq
and hence \eq{9-2} is equivalent to
\beql{9-4}
\Tr[E^{B}(\De)\bT\rh(t)]=\Tr[E^{B}(\De)\rh(t)].
\eeq
Let $\bT^{*}$ be the dual nonselective operation of $\bA(\ba)$.
It follows from \eq{9-4} that \eq{9-2} is equivalent to
\beql{9-5}
\Tr[(\bT^{*}E^{B}(\De))\rh(t)]=\Tr[E^{B}(\De)\rh(t)].
\eeq
Since $\rh(t)$ is arbitrary, \eq{9-2} is equivalent to
\beql{9-6}
\bT^{*}E^{B}(\De)=E^{B}(\De).
\eeq
Thus, we conclude that {\em the instantaneous measurement  using the 
apparatus $\bA(\ba)$ with nonselective operation $\bT$ does not 
disturb the observable $B$ if and only if \eq{9-6} holds for 
any Borel set $\De$.}

Now we are ready to state the answer to our problem.

{\bf Theorem 4.}
{\em Let $\bA(\ba)$ be an apparatus measuring an observable $A$
instantaneously and let $\bA(\bb)$ be an arbitrary apparatus
measuring an observable $B$.  Then, the successive measurement
using $\bA(\ba)$ and $\bA(\bb)$ is a simultaneous
measurement of $A$ and $B$ if and only if $\bA(\ba)$ does not
disturb $B$.
}

{\em Proof.}
It suffices to show the equivalence between \eq{12} and \eq{9-6}.
From \eq{T3'} and \eq{0414a}, we have
\begin{eqnarray*}
\lefteqn{\Pr\{\ba(t)\in\Delta,\bb(t+\Delta t)\in\Delta'\}}\\
&=&
{\rm Tr}[E^{B}(\Delta')\bX(\De)\rho(t)]\\
&=&
{\rm Tr}[E^{B}(\Delta')\bX(\R)[\rho(t)E^{A}(\De)]]\\
&=&
{\rm Tr}[[\bT^{*}E^{B}(\Delta')]\rho(t)E^{A}(\De)]\\
&=&
{\rm Tr}[E^{A}(\De)(\bT^{*}E^{B}(\Delta'))\rho(t)]
\end{eqnarray*}
Thus, the joint probability distribution of $A$ and $B$ 
is given by
\beql{1016-3}\label{eq:9-7}
\Pr\{\ba(t)\!\in\!\Delta,\bb(t\!+\!\Delta t)\!\in\!\Delta'\}
\!=\!
{\rm Tr}[E^{A}(\Delta)(\bT^{*}E^{B}(\De')\rho(t)].
\eeq
If (\ref{eq:9-6}) holds, (\ref{eq:12}) follows immediately
from (\ref{eq:9-7}).
Conversely, suppose that (\ref{eq:12}) holds.
By substituting $\Delta=\R$ in (\ref{eq:12}),
we have
\begin{equation}\label{eq:1016-4}
\Pr\{\ba(t)\in\R,\bb(t+\Delta t)\in\Delta'\}={\rm Tr}[E^{B}(\Delta')\rho(t)].
\end{equation}
On the other hand, from (\ref{eq:1016-3}) we have
\begin{eqnarray}\label{eq:1016-5} 
\lefteqn{\Pr\{\ba(t)\in\R,\bb(t+\Delta t)\in\Delta'\}}\qquad\nonumber\\
&=&
{\rm Tr}[(\bT^{*}E^{B}(\Delta'))\rho(t)].
\end{eqnarray}
Since $\rho(t)$ is arbitrary, from (\ref{eq:1016-4}) and (\ref{eq:1016-5}) we
obtain (\ref{eq:9-6}).
Therefore, (\ref{eq:12}) and (\ref{eq:9-6}) are equivalent.
\qed

From Theorems {1} and {4}, 
we can see that if the apparatus $\bA(\ba)$ 
measuring instantaneously
an observable $A$ does not disturb an observable $B$, then
$A$ and $B$ necessarily commute.
Therefore, we can conclude the following statement.

{\bf Theorem 5.}
{\em  Every apparatus measuring an observable disturbs 
all the observables that do not commute with the measured observable.
}

\subsection{Disturbance in indirect measurements}

By (\ref{eq:803a}) and by the property of the partial trace, we have
\begin{eqnarray*}
\lefteqn{{\rm Tr}[E^{B}(\Delta)\rho(t+\Delta t)]}\qquad\\
&=&{\rm Tr}\left[E^{B}(\Delta)
{\rm Tr}_{\cK}[U(\rho(t)\otimes\sigma)U^{\dagger}]\right]\\
&=&{\rm Tr}[(E^{B}(\Delta)\otimes I)U(\rho(t)\otimes\sigma)U^{\dagger}]\\
&=&{\rm Tr}[U^{\dagger}(E^{B}(\Delta)\otimes I)U
(I\otimes\sigma)(\rho(t)\otimes I)]\\
&=&{\rm Tr}\left[
{\rm Tr}_{\cK}[U^{\dagger}(E^{B}(\Delta)\otimes I)U(I\otimes\sigma)]
\rho(t)\right].
\end{eqnarray*}
Hence, \eq{10} is equivalent to 
\beqas
&\Tr\left[
\Tr_{\cK}[U^{\da}(E^{B}(\De)\otimes I)U(I\otimes\sigma)]\rh(t)
\right]&\\
&={\rm Tr}[E^{B}(\Delta)\rho(t)].&
\eeqas
Since $\rho(t)$ is arbitrary, (\ref{eq:10}) is equivalent to
\begin{equation}\label{eq:11}
{\rm Tr}_{{\cK}}[U^{\dagger}(E^{B}(\Delta)\otimes I)U(I\otimes\sigma)]
=E^{B}(\Delta)
\end{equation}
for any Borel set $\Delta$.

Obviously from (\ref{eq:11}), if $U$ and $B\otimes I$ commute, i.e., 
\begin{equation}\label{eq:110c}
[U,E^{B}(\Delta)\otimes I]=0
\end{equation}
for any Borel set $\Delta$,
then the $A$-measurement does not disturb the observable $B$.
However, (\ref{eq:110c}) is not a necessary condition for nondisturbing
measurement.  In the case where $\sigma$
is a pure state $\sigma=\ket{\xi}\bra{\xi}$, 
from (\ref{eq:11}) we have the following theorem.

{\bf Theorem 6.}
{\em Let $\bA(\ba)$ be an apparatus measuring an observable $A$ 
instantaneously
with indirect measurement model $(\cK,\ket{\Ph}\bra{\Ph},U,M)$.
The apparatus $\bA(\ba)$ does not disturb an observable $B$ 
if and only if
\begin{equation}\label{eq:110e}
[U,E^{B}(\Delta)\otimes I]\ket{\psi\otimes\Ph}=0
\end{equation}
for any Borel set $\Delta$ and any state vector $\psi$ of 
${\bf S}$.
}

{\em Proof.}
First, we note that in the case where $\si=\ket{\Ph}\bra{\Ph}$,
relation \eq{11} holds if and only if
\beq\label{eq:0422a}
\bracket{\ps\otimes\Ph|
U^{\dagger}(E^{B}(\Delta)\otimes I)U|\ps\otimes\Ph}
=\bracket{\ps|E^{B}(\Delta)|\ps}
\eeq
holds for any state vector $\ps$.
Suppose that \eq{110e} holds.
We have
$$
U(E^{B}(\De)\otimes I)\ket{\ps\otimes\Ph}
=(E^{B}(\De)\otimes I)U\ket{\ps\otimes\Ph}.
$$
Multiplying $U^{\da}$ from the left, we have
$$
(E^{B}(\De)\otimes I)\ket{\ps\otimes\Ph}
=U^{\da}(E^{B}(\De)\otimes I)U\ket{\ps\otimes\Ph},
$$
and hence, we have \eq{0422a}. 
Thus, if \eq{110e} holds 
for any Borel set $\Delta$ and any state vector $\psi$ 
then $\bA(\ba)$ does not disturb $B$.
Conversely, suppose that $\bA(\ba)$ does not disturb $B$.
Then, from \eq{11} with $\si=\ket{\Ph}\bra{\Ph}$, we have
\beqas
\lefteqn{\bracket{\ph'\otimes\Ph|
U^{\dagger}(E^{B}(\Delta)\otimes I)U|\ph\otimes\Ph}}\quad\\
&=&\bracket{\ph'\otimes\Ph|E^{B}(\Delta)\otimes I|\ph\otimes\Ph}
\eeqas
for any vectors $\ph,\ph'\in\cH$.
Let $\ps$ be a state vectors.
Put $\ket{\ph}=\ket{\ps}$ and $\ket{\ph'}=E^{B}(\De)\ket{\ps}$, 
we have
\beqa\label{eq:0422b}
&\bracket{\ps\otimes\Ph|
(E^{B}(\Delta)\otimes I)U^{\da}(E^{B}(\Delta)\otimes I)U
|\ps\otimes\Ph}&\nn\\
&=\bracket{\ps\otimes\Ph|E^{B}(\Delta)\otimes I|\ps\otimes\Ph}.&
\eeqa
By taking complex conjugate, we have
\beqa\label{eq:0422c}
&\bracket{\ps\otimes\Ph|
U^{\da}(E^{B}(\Delta)\otimes I)U(E^{B}(\Delta)\otimes I)
|\ps\otimes\Ph}&\nn\\
&=\bracket{\ps\otimes\Ph|E^{B}(\Delta)\otimes I|\ps\otimes\Ph}.&
\eeqa
From \eq{0422a} -- \eq{0422c}, we have
\beqas
\lefteqn{
\|\,[E^{B}(\De)\otimes I-U^{\da}(E^{B}(\De)\otimes I)U]
\ket{\ps\otimes\Ph}\|^{2}}\quad\\
&=&\bracket{\ps\otimes\Ph|E^{B}(\De)\otimes I|\ps\otimes\Ph}\\
& &\mb{}-\bracket{\ps\otimes\Ph|(E^{B}(\De)\otimes I)
                  U^{\da}(E^{B}(\De)\otimes I)U|\ps\otimes\Ph}\\
& &\mb{}-\bracket{\ps\otimes\Ph|U^{\da}(E^{B}(\De)\otimes I)U
                  (E^{B}(\De)\otimes I)|\ps\otimes\Ph}\\
& &\mb{}+\bracket{\ps\otimes\Ph|U^{\da}(E^{B}(\De)\otimes I)U
                  |\ps\otimes\Ph}\\
&=&0.
\eeqas
Thus, we have
$$
(E^{B}(\De)\otimes I)\ket{\ps\otimes\Ph}
=
U^{\da}(E^{B}(\De)\otimes I)U\ket{\ps\otimes\Ph}.
$$
Multiplying $U$ from the left, we have
$$
[U(E^{B}(\De)\otimes I)-
(E^{B}(\De)\otimes I)U]\ket{\ps\otimes\Ph}=0,
$$
and hence we have \eq{110e}.  Therefore, we conclude that if $\bA(\ba)$
does not disturb $B$, then \eq{110e} holds for any Borel set $\De$ and
any state vector $\ps$ of $\bS$.
\qed

\section{Local Measurements of Observables of Two Entangled System}

If the two observables to be measured belong to two 
different subsystems respectively, then they commute
each other and the measurement of one is not considered 
to disturb the other in general, 
so that the result obtained in the preceding
section applies to this situation.
The purpose of this section is to state this fact in the
rigorous language.

Let $C$ be an observable of an system $\bS_{1}$ with Hilbert space $\cH_{1}$
and $D$ an observable of another system $\bS_{2}$ with Hilbert space
$\cH_{2}$.
Suppose that the composite system $\bS=\bS_{1}+\bS_{2}$ is in the
state $\rh(t)$ at the time $t$.
Let us suppose that one measures the observable $C$ at the time $t$ using an 
apparatus $\bA(\ba)$ and that at the time, $t+\De t$,
just after the $C$-measurement one measures $D$ using any apparatus
$\bA(\bb)$ measuring $D$.
We assume that after the time $t$ there is no interaction between
$\bS_{1}$ and $\bS_{2}$.

First, we shall consider the case where the measurement of $C$ satisfies
the projection postulate.
In this case, in the composite system $\bS_{12}$ 
the observable $A=C\otimes I_{2}$ is measured at the time $t$ 
and the observable $B=I_{1}\otimes D$ is measured immediately after the
$A$-measurement, where $I_{1}$ and $I_{2}$ are the identity operators
on $\cH_{1}$ and $\cH_{2}$, respectively.  
From Theorem {2} the joint probability
distribution satisfies
\beqa\label{eq:723c}
\lefteqn{\Pr\{\ba(t)\in\De,\bb(t+\De t)\in\De'\}}\nn\\
&=&
\Tr\left[\left(E^{C}(\De)\otimes E^{D}(\De')\right)\rh(t)\right].
\eeqa

In order to compare this result with the argument given 
by EPR \cite{EPR35}, 
let us consider the special case where $C$ and $D$ are nondegenerate 
observables in their own subsystems and the initial state 
$\rh(t)$ is a pure state.
In this case, the state $\rh(t)$ is represented by a state
vector $\Ps(t)$ in the Hilbert space $\cH=\cH_{1}\otimes\cH_{2}$
as
$$
\rh(t)=\ket{\Ps(t)}\bra{\Ps(t)}.
$$
Let us suppose that the observables $C$ and $D$ have the spectral 
decompositions
\beqas
C&=&\sum_{n}a_{n}\ket{\ph_{n}}\bra{\ph_{n}},\\
D&=&\sum_{m}b_{m}\ket{\xi_{m}}\bra{\xi_{m}}.
\eeqas

EPR expand $\Ps(t)$ using the basis $\{\ph_{n}\}$ of $\cH_{1}$
as 
\beql{804f}
\Ps(t)=\sum_{n}\ket{\ph_{n}\otimes\et_{n}},
\eeq
where $\et_{n}$ are uniquely determined vectors in $\cH_{2}$
not necessarily orthogonal and, according to EPR,
are to be regarded merely as the coefficients of the expansion of
$\Ps(t)$ into a series of orthogonal vectors $\ph_{n}$.
Then, EPR considered the process of ``reduction of the
wave packet''
\beql{804h}
\sum_{n}\ket{\ph_{n}\otimes\et_{n}}
\mapsto
N\ket{\ph_{n}\otimes\et_{n}},
\eeq
where $N$ is the normalization constant determined up to a phase factor
by
\beq
N=\|\ph_{n}\otimes\et_{n}\|^{-1},
\eeq
and stated that
the state after the measurement conditional upon the
outcome $\ba(t)=a_{n}$ is determined as
\beql{0428a}
\Ps(t+\De t|\ba(t)=a_{n})
=
N\ket{\ph_{n}\otimes\et_{n}},
\eeq
where 
\beqa
\lefteqn{\ket{\Ps(t+\De t|\ba(t)=a_{n})}
         \bra{\Ps(t+\De t|\ba(t)=a_{n})}}\nn\quad\\
&=&\rh(t+\De t|\ba(t)=a_{n}).
\eeqa
From this, we have the joint probability formula
\beqa\label{eq:804e}
\lefteqn{\Pr\{\ba(t)=a_{n},\bb(t+\De t)=b_{m}\}}\qquad\nn\\
&=&|\braket{\ph_{n}\otimes\xi_{m}|\Ps(t)}|^{2},
\eeqa
which is a special case of \eq{723c}.

Now, let us show that the EPR argument is equivalent 
with the argument based on the projection postulate 
for the $A$-measurement.
From the projection postulate, if the outcome of the $\bA(\ba)$-measurement
is $a_{n}$, the state of the composite system at the time just after
the measurement is 
\beql{0428b}
\Ps(t+\De t|\ba(t)=a_{n})
=
\frac{(\ket{\ph_{n}}\bra{\ph_{n}}\otimes I_{2})\Ps(t)}
{\|(\ket{\ph_{n}}\bra{\ph_{n}}\otimes I_{2})\Ps(t)\|}.
\eeq
Then, from \eq{804f} we have
\beql{804g}
(\ket{\ph_{n}}\bra{\ph_{n}}\otimes I_{2})\Ps(t)
=\ket{\ph_{n}\otimes\et_{n}}.
\eeq
Thus, we have shown that \eq{0428a} is the consequence 
from the projection postulate \eq{0428b}.

In the following, we shall consider the general case.
For instance, consider the case where the $A$-measurement leaves 
the system $\bS_{1}$ in a fixed state $\ph_{1}$ independent of
the outcome such as the vacuum state after photon counting.
Does \eq{723c} hold even in this case? 
The answer to this question might depend on the way of measuring $A$.
However, if the measurement of $A$ is carried out so as  not to affect
the system $\bS_{2}$, then from the result in the preceding section
we will be able to conclude relation \eq{723c}.
In order to ensure that the measurement of $A$ does not affect
the system $\bS_{2}$, we introduce the following condition.

We will say that the apparatus $\AA$ measuring $A$
is local in the system $\bS_{1}$ if the measuring interaction
is confined in the system $\bS_{1}$ and the apparatus $\bA(\ba)$,
as formulated precisely as follows.
Let $\cK$ be the Hilbert space of the probe $\bP$ in the apparatus
$\AA$ and suppose that $\bP$ is prepared in the state $\si$ at the time
$t$ of measurement and let $U$ be the unitary operator of 
$\cK\otimes\cH_{1}\otimes\cH_{2}$ representing the time evolution
of the composite system $\bS+\bP$.  
Then, the apparatus $\AA$ is said to be {\em local} 
in the system $\bS_{1}$ if we have
\begin{equation}\label{eq:110d}
[U,I_{1}\otimes X\otimes I_{\cK}]=0,
\end{equation}
for any bounded operator $X$ on ${\cal H}_{2}$, 
where $I_{\cK}$ is the identity on $\cK$.

{\bf Theorem 7.}
{\em Suppose that the composite system $\bS=\bS_{1}+\bS_{2}$ is in
the state $\rh(t)$ at the time $t$ of measurement.
Let $C$ and $D$ be observables of $\bS_{1}$ and $\bS_{2}$, respectively.
If the apparatus $\AA$ measuring $A=C\otimes I_{2}$ instantaneously 
is local in the system $\bS_{1}$ then {\rm \eq{723c}} holds.
}

{\em Proof.}
Let $\si$ be the state of the probe at $t$.
From Theorem {5} it suffices to show that $\bA(\ba)$ 
does not disturb the observable $B=I\otimes D$.
By assumption, we have 
\beqas
[U,E^{B}(\De)\otimes I_{\cK}]
&=&[U,I_{1}\otimes E^{D}(\De)\otimes I_{\cK}]\\
&=&0
\eeqas
for all Borel set $\De$.  Thus, relation \eq{110c} holds,
so that $\bA(\ba)$ does not disturb the observable $B=I_{1}\otimes D$.
Therefore, equation \eq{723c} follows from Theorem {4}.
\qed

From the above theorem, we have also the following statement: 
{\em Any pair of local instantaneous measuring apparatuses 
of $A=C\otimes I_{2}$ and $B=I_{1}\otimes D$ satisfies 
the joint probability formula 
\begin{equation}
\Pr\{\ba(t)\in\Delta,\bb(t)\in\Delta'\}
={\rm Tr}[(E^{C}(\Delta)\otimes E^{D}(\Delta'))\rho(t)],
\end{equation}
regardless of the order of the measurement, where we identify
$t$ with $t+\De t$.}

In the EPR paper \cite{EPR35}, the so called EPR correlation
is derived theoretically under the assumption that the pair of 
measurements satisfies the projection postulate, but the present 
result concludes that the EPR correlation holds for any pair of
local instantaneous measurements as experiments have already suggested.

\section{Minimum disturbing measurements}

Classical measurements are usually considered to disturb no
measured systems.  This does not mean, however, that no classical 
measurement disturbs the system but means that
among all the possible measurement the minimum disturbing 
measurement does not disturb the system in principle.
In this section, we shall introduce the notion of the minimum
disturbing measurement in quantum mechanics and show that
this is equivalent to the measurement satisfying the projection
postulate.

For an apparatus $\bA(\bx)$, we denote by $\cD(\bx)$ the set
of observables that are disturbed by $\bA(\bx)$, i.e., 
$\cD(\bx)$ is the set of observables $B$ such that 
$\bT^{*}E^{B}(\De)\not=E^{B}(\De)$ for some Borel set $\De$,
where $\bT$ is the nonselective operation of $\bA(\bx)$.
Let $A$ be an observable of the system $\bS$ and 
let $\bA(\ba)$ be an apparatus measuring $A$ instantaneously.
The apparatus $\bA(\ba)$ is called {\em minimum disturbing} if 
$\cD(\ba)\subset\cD(\bx)$ for any apparatus $\bA(\bx)$ 
measuring $A$ instantaneously.
Then, we have the following statement. 

{\bf Theorem 8.}
{\em Let $\bA(\ba)$ be an apparatus measuring a discrete observable $A$
instantaneously.  The apparatus $\bA(\ba)$ is minimum disturbing 
if and only if $\bA(\ba)$ satisfies the projection postulate.
}

{\em Proof.}
Let $\cC(A)$ be the set of observables that do not commute with $A$.
From Theorem {5}, we have 
\beql{0425a}
\cC(A)^{c}\subset\cD(\bx)
\eeq
for any apparatus $\bA(\bx)$ measuring $A$ instantaneously, 
where ${}^{c}$ stands for the complement in the set of observables.
Let $\bA(\ba)$ be an apparatus measuring $A$ instantaneously.
Suppose that $\bA(\ba)$ satisfies the projection postulate.
Then, from Theorem {2} we have
\beql{0425b}
\cD(\ba)\subset\cC(A)^{c},
\eeq
and hence from \eq{0425a} we conclude that $\bA(\ba)$ is minimum
disturbing and 
\beql{0425c}
\cD(\ba)=\cC(A)^{c}.
\eeq
Conversely, suppose that $\bA(\ba)$ is minimum disturbing.
We have an indirect measurement model that measures
$A$ instantaneously and satisfying the projection postulate 
\cite{84QC}.
Hence, there is an apparatus $\bA(\bx)$ measuring
$A$ instantaneously such that $\cD(\bx)=\cC(A)^{c}$.
By assumption, $\bA(\ba)$ is minimum disturbing, so that 
$\cD(\ba)=\cC(A)^{c}$.  
Then, the operation valued measure $\bX$ of $\bA(\ba)$ is such that
$\bX(\R)^{*}E^{B}(\De')=E^{B}(\De')$ 
for all $B\in\cC(A)$ and $\De'\in\cB(\R)$.
Thus, we have
\beqas
\lefteqn{\Tr[E^{B}(\De')\bX(\De)\rh(t)]}\quad\\
&=&\Tr[[\bX(\De)^{*}E^{B}(\De')]\rh(t)]\\
&=&\sum_{a\in\De}\Tr[[\bX\{a\}^{*}E^{B}(\De')]\rh(t)]\\
&=&\sum_{a\in\De}\Tr[(E^{A}\{a\}[\bX(\R)^{*}E^{B}(\De')]E^{A}\{a\})\rh(t)]\\
&=&\sum_{a\in\De}\Tr[E^{A}\{a\}E^{B}(\De')E^{A}\{a\}\rh(t)]\\
&=&\Tr[E^{B}(\De')\sum_{a\in\De}E^{A}\{a\}\rh(t)E^{A}\{a\}].
\eeqas
Since $B$ and $\De'$ are arbitrary, we have
$$
\bX(\De)\rh(t)=\sum_{a\in\De}E^{A}\{a\}\rh(t)E^{A}\{a\}
$$
and hence 
$$
\rh(t+\De t|\ba\in\De)=\frac{\sum_{a\in\De}E^{A}\{a\}\rh(t)E^{A}\{a\}}
{\Tr[E^{A}(\De)\rh]}.
$$
Thus, $\bA(\ba)$ satisfies the projection postulate.
\qed

We refer to \cite{84QC,DL70} for different approaches to the minimum 
disturbance condition.  The present approach leads to the simplest
characterization of the measurements satisfying the projection postulate,
which can be called eventually as the {\em minimum disturbing 
measurements}.

\section{Concluding remarks}

As anticipated from the ordinary interpretation of the uncertainty
principle or the principle of complementarity formulated by
the noncommutativity of observables,
every measurement of an observable disturbs every observable that 
does not commute with the measured observable.
It should be noticed, however, that this does not imply a prevailing 
interpretation of the Heisenberg uncertainty principle that 
the measurement of the position with accuracy $\ep$ must bring 
about an indeterminacy $\et=\hbar/2\ep$ in the value of the
position \cite[p.~239]{vN55}.
In fact, we can construct an indirect measurement model of the
postion measurement that counters the above statement \cite{98CQSR};
this model has the complete accuracy $\ep=0$ but disturbs the momentun 
arbitrarily small if the input state is arbitarily close to 
the momentum eigenstate.
This example suggests that the relation between the accuracy and
the disturbance is more complicated than the relation
$\ep\et\ge\hbar/2$ suggested by the Robertson uncertainty 
relation \cite{Rob29}.
The detailed investigation will be presented in the forthcomming 
article.



\begin{references}

\bibitem{vN55}
J. von~Neumann,
{\it Mathematical Foundations of Quantum Mechanics}
(Princeton University, Princeton, NJ, 1955).

\bibitem{Lud51}
G. {L\"{u}ders},
Ann.\ Phy.\ (Leipzig) (6) {\bf 8}, 322 (1951).

\bibitem{84QC}
M. Ozawa,
J. Math.\ Phys.\ {\bf 25}, 79 (1984).

\bibitem{85CA}
M. Ozawa,
Publ.\ Res.\ Inst.\ Math.\ Sci., Kyoto Univ.\ {\bf 21}, 279 (1985).

\bibitem{EPR35}
A. Einstein, B. Podolsky and N. Rosen,
Phys.\ Rev.\ {\bf 47}, 777 (1935).

\bibitem{AGR82}
A. Aspect, P. Grangier, and G. Roger,
Phys.\ Rev.\ Lett.\ {\bf 49}, 91 (1982).

\bibitem{ADR82}
A. Aspect, J. Dalibard, and G. Roger,
Phys.\ Rev.\ Lett.\ {\bf 49}, 1804 (1982).

\bibitem{BPMEWZ97}
D. Bouwmeester, J.-W. Pan, K. Mattle, M. Eibl, H. Weinfurter, and A. Zeilinger,
Nature (London) {\bf 390}, 575 (1997).

\bibitem{BBMHP97}
D. Boschi, S. Branca, F.~De Martini, L. Hardy, and S. Popescu,
Phys.\ Rev.\ Lett.\ {\bf 80}, 1121 (1998).

\bibitem{BD00}
C.~H. Bennett and D. DiVincenzo,
Nature (London) {\bf 404}, 247 (2000).

\bibitem{Dav76}
E.~B. Davies,
{\it Quantum Theory of Open Systems}
(Academic Press, London, 1976).

\bibitem{Kra83}
K. Kraus,
{\it States, Effects, and Operations: Fundamental Notions of Quantum
  Theory},
{\it Lecture Notes in Physics {\bf 190}} (Springer, Berlin, 1983).

\bibitem{85CC}
M. Ozawa,
J. Math.\ Phys.\ {\bf 26}, 1948 (1985).

\bibitem{86IQ}
M. Ozawa,
J. Math.\ Phys.\ {\bf 27}, 759 (1986).

\bibitem{89RS}
M. Ozawa,
in {\it Squeezed and Nonclassical Light}, 
edited by P. Tombesi and E.~R. Pike (Plenum, New York, 1989),
p.~263.


\bibitem{BLM91}
P. Busch, P.~J. Lahti, and P. Mittelstaedt,
{\it The Quantum Theory of Measurement},
{\it Lecture Notes in Physics {\bf m2}} (Springer, Berlin, 1991).

\bibitem{HK64}
R. Haag and D. Kastler,
J. Math.\ Phys.\ {\bf 5}, 848 (1964).

\bibitem{Lud67}
G. Ludwig,
Commun.\ Math.\ Phys.\ {\bf 4}, 331 (1967).

\bibitem{DL70}
E.~B. Davies and J.~T. Lewis,
Commun.\ Math.\ Phys.\ {\bf 17}, 239 (1970).

\bibitem{83CR}
M. Ozawa,
in {\it Probability Theory and Mathematical Statistics, 
Lecture Notes in Mathematics {\bf 1021}},
edited by K. It\^{o} and J.~V. Prohorov 
(Springer, Berlin, 1983), p.~518.

\bibitem{00MN}
M. Ozawa,
Measurements of nondegenerate discrete observables,
Phys.\ Rev.\ A (to appear),
[Online preprint: LANL quant-ph/0003033].

\bibitem{Dir58}
P.~A.~M. Dirac,
{\it The Principles of Quantum Mechanics}, 
4th ed.\ (Oxford University, Oxford, 1958).

\bibitem{Sch35}
E. {Schr\"{o}dinger},
Naturwissenshaften {\bf 23}, 807, 823, 844 (1935)
[Proc.\ Am.\ Philos.\ Soc.\ {\bf 124}, 323 (1980)].

\bibitem{98PT}
M. Ozawa,
in {\it Fifth International Conference on Squeezed States and 
Uncertainty Relations}, 
edited by D. Han, J. Jansky, Y.~S. Kim, and V.~I. Man'ko
 (NASA, Goddard, 1998), p.~517.

\bibitem{SD81}
M.~D. Srinivas and E.~B. Davies,
{\it Optica Acta} {\bf 28}, 981 (1981).

\bibitem{IUO90}
N. Imoto, M. Ueda, and T. Ogawa,
Phys.\ Rev.\ A {\bf 41}, 4127 (1990).

\bibitem{88MS}
M. Ozawa,
Phys.\ Rev.\ Lett.\ {\bf 60}, 385 (1988).

\bibitem{90QP}
M. Ozawa,
Phys.\ Rev.\ A {\bf 41}, 1735 (1990).


\bibitem{Mad88}
J. Maddox,
Nature (London) {\bf 331}, 559 (1988).

\bibitem{LB39}
F. London and E. Bauer,
{\it La th\'{e}orie de l'observation en m\'{e}canique quantique}
(Hermann, Paris 1939) 
[in {\it Quantum Theory and Measurement}, 
edited by J.~A. Wheeler and W.~H. Zurek
 (Princeton University, Princeton, NJ, 1983), p~217].

\bibitem{Wig63}
E.~P. Wigner,
Am.\ J. Phys.\ {\bf 31}, 6 (1963).

\bibitem{97OQ}
M. Ozawa,
Ann.\ Phys.\ (N.Y.) {\bf 259}, 121 (1997).

\bibitem{00MO}
M. Ozawa (in preparation).

\bibitem{95MM}
M. Ozawa,
in  {\it Quantum Communications and Measurement},
edited by V.~P. Belavkin, O. Hirota, and R.~L. Hudson 
(Plenum, New York, 1995), p.~109.

\bibitem{93CA}
M. Ozawa,
J. Math.\ Phys.\ {\bf 34}, 5596 (1993).

\bibitem{98CQSR}
M. Ozawa, Controlling Quantum State Reduction,
Online preprint: LANL quant-ph/9805033.

\bibitem{Rob29}
H. P. Robertson,
Phys.\ Rev.\ {\bf 34}, 163 (1929).

\end{references}
\end{document}